\documentclass[%
 reprint,
 amsmath,amssymb,
 aps,
]{revtex4-2}

\usepackage{graphicx}
\usepackage{dcolumn}
\usepackage{bm}

\usepackage{amsmath,amssymb}
\usepackage{array}
\usepackage{multirow}
\usepackage{graphicx}
\usepackage{hhline}
\usepackage[cdot,mediumqspace,amssymb]{SIunits}
\usepackage{xfrac}
\usepackage[normalem]{ulem}

\newcommand{\vect}[1]{\boldsymbol{#1}}

\begin{document}
\preprint{APS/123-QED}

\title{Vacuum polarization in molecules II: higher order corrections}

\author{D. J. Flynn}
\affiliation{School of Physics, The University of Melbourne, Victoria 3010, Australia}

\author{I. P. Grant}
\affiliation{Mathematical Institute, University of Oxford, Oxford, United Kingdom}
\affiliation{Department of Applied Mathematics and Theoretical Physics, University of Cambridge, Cambridge, United Kingdom}

\author{H. M. Quiney}
\affiliation{School of Physics, The University of Melbourne, Victoria 3010, Australia}

\date{\today}

\begin{abstract}
We outline a general formalism for treating vacuum polarization phenomena within an effective field expansion. The coupling between source charges and virtual fields is examined from the perspectives of electrostatic potentials, induced charge densities and form factors in momentum space. A strategy for the efficient calculation of vacuum polarization potentials is outlined, implemented, and applied towards the construction of fitting potentials that are suitable for molecular electronic structure calculations, which enclose no overall charge by construction. The order $\alpha(Z \alpha)$, $\alpha (Z \alpha)^{3}$ and $\alpha^{2}(Z\alpha)$ effects of a Gaussian nuclear charge on the electron-positron field are applied variationally towards the E119F molecule, as well as the order $\alpha(Z \alpha)$ effects arising from the virtual muon and charged pion fields.
\end{abstract}

\maketitle


\section{Introduction}\label{sec:introduction}
In addition to studies in spectroscopy, chemistry and condensed matter physics, atoms and molecules are used to test quantum electrodynamics and electroweak theory, and in searches for new physics beyond the Standard Model of particle physics. They are able to fulfil this role because of the strong electrostatic field in the neighbourhood of heavy nuclei. Analysis of these experimental applications requires, however, a detailed knowledge of the electronic wavefunctions in this region. This approach then demands a relativistic treatment of electronic amplitudes, detailed models of nuclear charge-current distributions, the inclusion of Breit corrections to the Coulomb interactions between electrons, many-body corrections to the wavefunction, and the incorporation of quantum electrodynamical (QED) effects, such as the electron self-energy and vacuum polarization. Electroweak interactions also require the inclusion of additional couplings between the electronic and nuclear degrees of freedom. The treatment of many-body corrections is formally simpler if as much of the underlying physics as possible can be absorbed within a one-body mean-field potential that is used to define a complete set of single-electron amplitudes. Given these complications, the choice of computational techniques and the judicious use of effective approximations require close attention.\\

Chief amongst the challenges of interfacing QED with relativistic molecular electronic structure calculations is the apparent complexity of the electrostatic potentials involved. We have recently discussed the evaluation of higher-order vacuum polarization corrections from first principles, by direct evaluation of external-field Feynman diagrams in a Gaussian basis set~\cite{grant2022grasp}. The need to observe explicitly the charge conjugation symmetry places strong demands on the basis set that is required, which is a work in progress in our molecular electronic structure code, \texttt{BERTHA}. Vacuum polarization effects give rise, however, to local effective potentials that are algebraically complicated but otherwise straightforward to evaluate in an order-by-order perturbative approach. The accuracy of such calculations is largely determined by the accuracy of the electronic charge density in the neighbourhood of the nuclei, which can be ensured using conventional basis set constructions of modest dimension. Our {\em ab initio} approach to vacuum polarization~\cite{grant2022grasp} has the advantage of summing some classes of Feynman diagram through all orders in perturbation theory, but the disadvantage that very large basis sets are required to achieve convergence. These differences in behaviour arise because in the {\em ab initio} approach, it is necessary to construct both the charge density and the effective interaction in a basis set representation.\\

Within the sub-critical field regime, one is bound only by the assumption that there are no local fields of sufficient strength to trigger on-shell pair production, as indeed is the case for point-nuclear atomic systems with $Z \lesssim 137$. This permits the textbook expansion technique~\cite{ItzyksonZuber} which makes available a number of electrostatic effective potentials by means of free-field formulations based in scattering theory. Those potentials comprise the theoretical building blocks of the current work, where we focus on atoms and molecules for which the expansion parameter $Z \alpha$ approaches unity. One is then required either to develop all-order {\em ab initio} methods or to carry the expansion of vacuum polarization effects beyond the leading-order Uehling potential around an atomic nucleus. In the first instance, we pursue the latter approach in this work.\\

In preliminary work~\cite{FlynnQuiney2024a}, we implemented computational methods for the evaluation of the leading-order Uehling interaction in relativistic electronic structure calculations. The approach that we have taken is to expand the effective potential in an auxiliary Gaussian basis that strictly constrains the net induced vacuum charge to vanish identically. This approach facilitates the subsequent evaluation of interaction matrix elements in relativistic mean-field models. Here, we extend this work to include higher-order quantum electrodynamical effects, and polarization phenomena involving virtual fields other than the electron-positron field.\\

In Sec.~\ref{sec:formalism}, we develop a generalized formalism that facilitates the incorporation of Uehling, K\"all\'en-Sabry and Wichmann-Kroll interactions within relativistic molecular electronic structure models. The technical details of the computational implementation of this approach are presented in Sec.~\ref{sec:implementation}. Calculations of these effects in the atomic species E119$^+$ and the E119F molecule are presented in Sec.~\ref{sec:results-discussion} using a selection of model hamiltonians. Conclusions from these studies and perspectives for future work are discussed in Sec.~\ref{sec:conclusion}.

\section{Formalism}\label{sec:formalism}
In conventional formulations of quantum electrodynamics, bound-state vacuum polarization phenomena are calculated by invoking an $S$-matrix expansion in a free-particle basis set. This leads to the characterization of these phenomena as a series of terms involving the perturbative expansion variables $\alpha^m(Z\alpha)^n$, where $m=1,2,\ldots$, $n=1,3,5,\ldots$, $Z$ represents the nuclear charge, and $\alpha$ is the fine-structure constant. The Uehling, Wichmann-Kroll and K\"all\'{e}n-Sabry potentials provide the leading contributions which are, respectively, of order $\alpha(Z\alpha)$, $\alpha(Z\alpha)^3$ and $\alpha^2(Z\alpha)$.\\

A knowledge of the vacuum polarization potentials generated by the external Coulomb field of a point charge may be used to construct the corresponding quantities produced by more detailed distributed models of nuclear structure. In this section, we establish the detailed connections between renormalized vacuum charge densities, electrostatic potentials, spectral functions and form factors represented in Fig.~\ref{fig:vp-connections}. In practice, these connections allow us to formulate convenient tools for the calculation of vacuum polarization effects that are consistent with existing computational approaches to atomic and molecular physics.

\subsection{Momentum space representations}\label{subsec:momentum-space}

\begin{figure}[!b]
    \begin{center}
    \includegraphics[width=8.5cm]{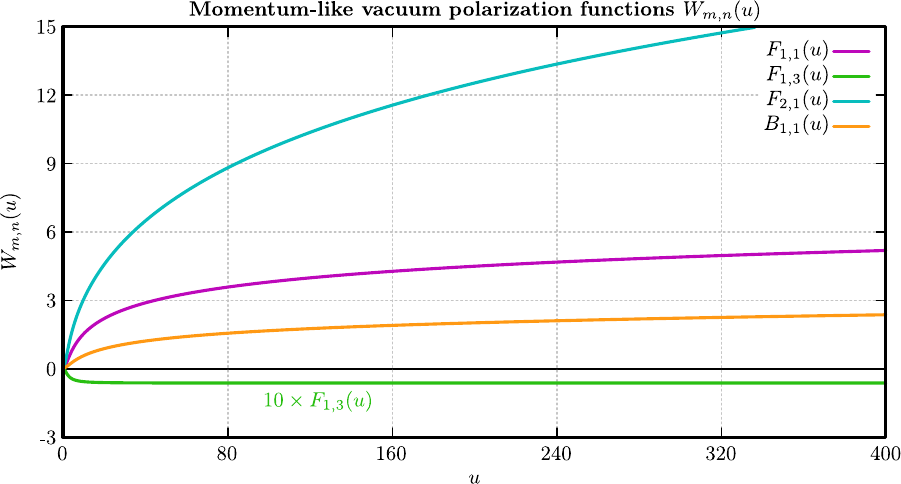}
    \includegraphics[width=8.5cm]{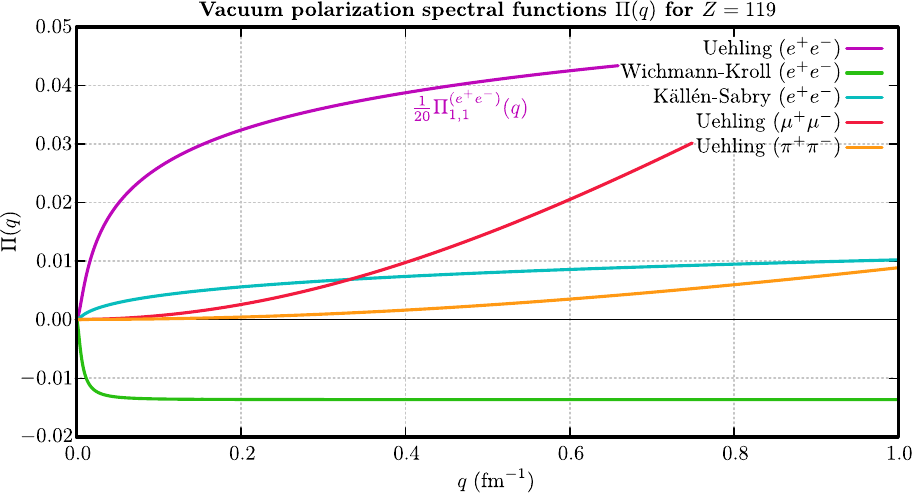}
    \caption{Momentum space functions $W_{m,n}(u)$ that generate the vacuum polarization spectral functions, $\Pi_{m,n}^{w^{+}w^{-}}(q)$. Some quantities have been scaled for presentation.}
    \label{fig:momentum-like-functions}
    \end{center}
\end{figure}

The coupling between an electrostatic source and a charge-carrying virtual field may be represented by momentum space quantities of the form
\begin{equation}\label{eq:spectral-function}
\Pi_{m,n}(\lambdabar q) = \bigl( \tfrac{\alpha}{\pi} \bigr)^{m} (Z \alpha)^{n} \tfrac{1}{\alpha} W_{m, n} (\lambdabar q).
\end{equation}\\
The Compton wavelength $\lambdabar = 1/mc$ depends on the mass, $m$, of the virtual quanta. It imposes a natural length scale of the coupling. In atomic units the electron-positron field couplings involve the electron mass $m_{e}=1$, and therefore any vacuum modifications that arise are localized within a radius of $\lambdabar_{e} = \alpha \approx 386 \; \text{fm}$ from their source. The functions $W_{m,n}(u)$ are dimensionless, as is the argument $u$. Their detailed forms depend on the nature of the virtual field from which they are constructed. These may be fermions or bosons, which we differentiate by the functions $F(u)$ and $B(u)$, respectively. For notational convenience, we often write $\Pi_{m,n}^{w^{+}w^{-}}(q)$ and infer the Compton wavelength $\lambdabar_{w}$ from relevant chared virtual field, such as in Fig.~\ref{fig:momentum-like-functions}.\\

\begin{widetext}
	\begin{center}
		\begin{figure}[!b]
			\includegraphics[width=14.0cm]{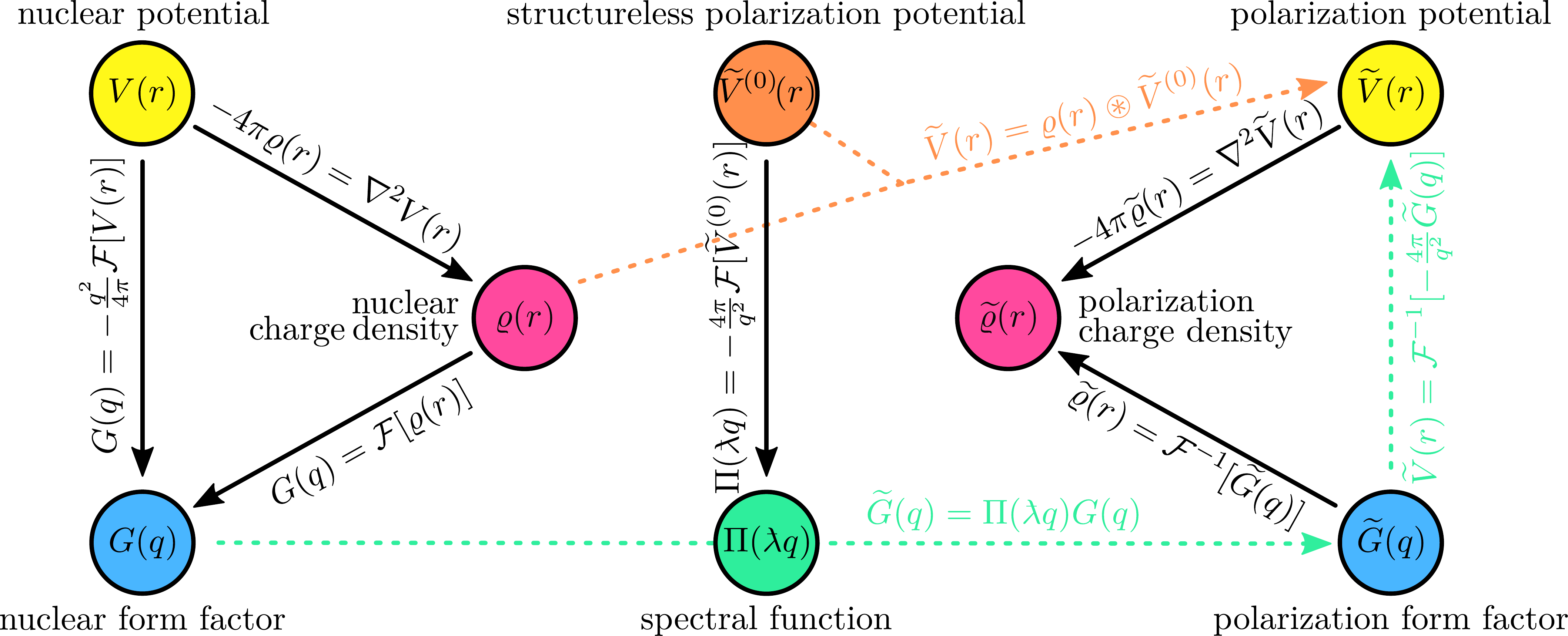}
			\caption{Connections between different quantities in our study of vacuum polarization. Rather than calculating polarization potentials by convolution (orange path), we make use of spectral functions and Fourier transforms (green path).}
			\label{fig:vp-connections}
		\end{figure}
	\end{center}
\end{widetext}

The spectral functions in Eq.~(\ref{eq:spectral-function}) measure the response of a charge-carrying virtual field due to the presence of a point-like or ``structureless'' source charge. While an atomic nucleus is not a point-like source, its finite charge structure is readily accommodated within this prescription by utilising its form factor, $G(q)$. The form factor is defined to be the Fourier transform of the charge density, $\varrho(\vect{r})$. If the charge density is spherically symmetric such that $\varrho(\vect{r}) = \varrho(r)$, then its form factor $G(\vect{q})=G(q)$ is calculated by
\begin{equation}\label{eq:source-form-factor}
G(q) = \int_{0}^{\infty} 4 \pi r^{2} \varrho(r) j_{0}(qr) \; \mathrm{d}r.
\end{equation}
where $j_n(x)$ are spherical Bessel functions~\cite{abramowitz1970}.\\

It is convenient to assume that the source charge distribution is normalized to unity, so that the parameter $Z$ appears as an explicit expansion parameter. The vacuum polarization form factor for a finite distribution of source charge is then given by 
\begin{equation}\label{eq:polarization-formfactor}
\widetilde{G}_{m,n}(q) = G(q) \Pi_{m,n}(\lambdabar q).
\end{equation}
On the assumption that the source charge $\varrho(r)$ is continuous and smooth, the long-range behaviour of the vacuum polarization form factor is attenuated by the source form factor, which satisfies $G(q) |_{q \to \infty} = 0 $. This attenuates the induced modification of the virtual field in such a way that unphysical cusp behaviour is eliminated. This is an important feature if these methods are to be applied to Gaussian basis set expansions of the electric charge-current density, which is the conventional approach in relativistic quantum chemistry.\\

The normalization condition on the source charge density requires that $G(0) = 1$. In the low momentum transfer limit (where $q$ is small), one may perform a Taylor expansion of the spherical Bessel function, obtaining
\begin{eqnarray}\label{eq:formfactor-powerseries}
G(q) 
 & \simeq & \sum_{k=0}^{\infty} \tfrac{(-1)^{k}}{(2k+1)!} q^{2k } \langle r^{2k} \rangle \\ \notag
 &=& 1 - \tfrac{1}{6} q^{2} \langle r^{2} \rangle + \tfrac{1}{120} q^{4} \langle r^{4} \rangle - \tfrac{1}{5040} q^{6} \langle r^{6} \rangle + \cdots.
\end{eqnarray}
The even radial charge moments of the source charge, $\langle r^{2k} \rangle$, significantly influence electronic structure calculations. As a result, a nuclear charge model must be parametrized to produce both the correct total charge, $Z$, and the appropriate nuclear root-mean-square radius, $R = \langle r^{2} \rangle^{1/2}$. A similar parameterization approach has been adopted here to model each of the vacuum polarization interaction terms.

\subsection{Electrostatic potentials}\label{subsec:potentials}

For a point-source charge, polarization potentials are
\begin{equation}\label{eq:vp-point-potential}
\widetilde{V}_{m,n}^{(0)} (r) 
 = - \bigl( \tfrac{\alpha}{\pi} \bigr)^{m} (Z \alpha)^{n} \tfrac{1}{\alpha r} \chi_{m,n} \bigl( \tfrac{2r}{\lambdabar} ; 1\bigr).
\end{equation}
\begin{figure}[!b]
    \begin{center}
    \includegraphics[width=8.5cm]{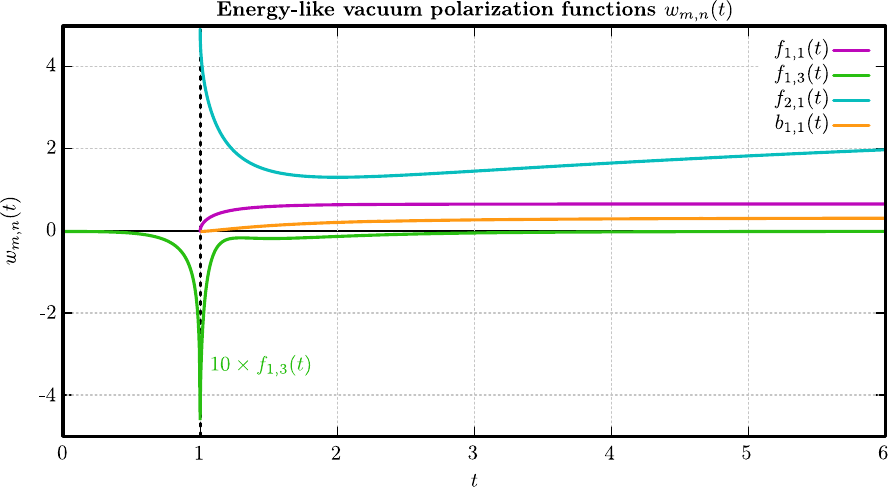}
    \caption{Energy-like unitless functions $w_{m,n}(t)$ required to generate vacuum polarization potentials. The Wichmann-Kroll function $f_{1,3}(t)$ has been modified for presentation.}
    \label{fig:wichmannkroll-ft}
    \end{center}
\end{figure}
This involves the auxiliary functions
\begin{equation}\label{eq:chi-functions}
\chi_{m,n}\bigl(x ; k \bigr) = \int_{\beta}^{\infty} t^{-k} w_{m,n}(t) \; e^{-xt} \; \mathrm{d}t,
\end{equation}
for which $\beta = 0$ or $\beta = 1$. For notational convenience we suppress the parameter $\beta$; it should be assumed that $\beta = 1$ unless another value is stated explicitly. The functions $w_{m,n}(t)$, shown in Fig.~\ref{fig:wichmannkroll-ft}, embody the features of a given vacuum polarization diagram, while the exponential behaviour imposes useful properties about the total charge embedded within a vacuum fluctuation. We distingish between fermion and boson fields in later sections by replacing $w_{m,n}(t)$ with $f_{m,n}(t)$ and $b_{m,n}(t)$, respectively. The variable $t$ may be regarded as an energy-like parameter.\\

The polarization potential arising from a finite source charge density $\varrho(\vect{r})$ requires that it be folded with the structureless potential using the convolution
\begin{equation*}
\widetilde{V}_{m,n} (\vect{r})
 = \varrho(\vect{r})\circledast \widetilde{V}_{m,n}^{(0)} (\vect{r})
 = \iiint \varrho(\vect{s}) \; \widetilde{V}_{m,n}^{(0)} (\vect{r}-\vect{s}) \; \mathrm{d}^{3}\vect{s}.
\end{equation*}\label{eq:vp-potential-finite}
For a spherically-symmetric source charge, the convolution can readily be be reduced to
\begin{eqnarray}\label{eq:vp-potentials-finite}
\lefteqn{\widetilde{V}_{m,n} (r)
 = -\bigl( \tfrac{\alpha}{\pi} \bigr)^{m} (Z \alpha)^{n} \tfrac{\pi \lambdabar}{\alpha r} \cdot }\\ \notag
 && \int_{0}^{\infty} s \varrho(s) \Bigl[ \chi_{m,n} \bigl(\tfrac{2|r-s|}{\lambdabar} ; 2 \bigr) - \chi_{m,n} \bigl( \tfrac{2(r+s)}{\lambdabar} ; 2 \bigr) \Bigr] \; \mathrm{d}s.
\end{eqnarray}
Alternatively, application of the Fourier differentiation theorem leads to   
\begin{eqnarray}\label{eq:polarization-potential-transform}
\widetilde{V}_{m,n}(r)
 & =& \mathcal{F}^{-1} \bigl[ - \tfrac{4 \pi}{q^{2}} G(q) \Pi_{m,n} (\lambdabar q) \bigr] \\ \notag
 &=& -\tfrac{2}{\pi} \int_{0}^{\infty} G(q) \Pi_{m,n} (\lambdabar q) j_{0}(rq) \; \mathrm{d}q.
\end{eqnarray}
The expression in Eq.~(\ref{eq:vp-potentials-finite}) can be compared against the form factor representation of Eq.~(\ref{eq:polarization-potential-transform}) to explicitly define the relationship between the pair of quantities $W(u)$ and $w(t)$. By taking the Fourier transform of each equation and replacing the spectral function with Eq.~(\ref{eq:spectral-function}), we find
\begin{widetext}
    \begin{equation}\label{eq:polarisation-formfactor-equation}
            G(q) W_{m, n} (\lambdabar q)
 = \tfrac{1}{4} \lambdabar^{2} q^{2} \int_{0}^{\infty} \tfrac{r}{\lambdabar} \biggl[ \int_{0}^{\infty} 4 \pi s \varrho(s) \biggl\{ \int_{\beta}^{\infty} \tfrac{1}{t^{2}} w_{m,n}(t)  \Bigl( e^{-2t|r-s|/\lambdabar} - e^{-2t(r+s)/\lambdabar} \Bigr) \; \mathrm{d}t \biggr\} \; \mathrm{d}s \biggr] j_{0}(rq) \; \mathrm{d}r.
    \end{equation}
\end{widetext}
By first performing the integral over the radial coordinate $r$, and then isolating an integral over the source charge radial coordinate $s$ to reflect $G(q)$, one obtains from Eq.~(\ref{eq:polarisation-formfactor-equation}) the transform relation
\begin{equation}\label{eq:transform-pair}
W_{m,n}(u) = - \int_{\beta}^{\infty} \tfrac{1}{t} \tfrac{u^{2}}{u^{2}+4t^{2}} w_{m,n}(t) \; \mathrm{d}t.
\end{equation}
The fermion function $W_{m,n}(u)=F_{m,n}(u)$ is generated by choosing $w_{m,n}(t)=f_{m,n}(t)$, while the boson function $W_{m,n}(u)=B_{m,n}(u)$ is generated with $w_{m,n}(t)=b_{m,n}(t)$.\\

There are limiting values of a vacuum polarization potential that are useful in establishing its asymptotic behaviour. The first of these involves the limit $r \to 0$, from which we can construct two equivalent representations for the potential at the origin:
\begin{eqnarray*}
\widetilde{V}_{m,n} (0)
 &=& - \bigl( \tfrac{\alpha}{\pi} \bigr)^{m} (Z \alpha)^{n} \tfrac{4 \pi}{\alpha} \int_{0}^{\infty} s \varrho(s) \; \chi_{m,n} \bigl(\tfrac{2s}{\lambdabar} ; 1\bigr) \; \mathrm{d}s\\
 &=& - \bigl( \tfrac{\alpha}{\pi} \bigr)^{m} (Z \alpha)^{n} \tfrac{2}{\pi \alpha} \int_{0}^{\infty} G(q) W_{m,n} (\lambdabar q) \; \mathrm{d}q.
\end{eqnarray*}

At large radial distances $r$, far from the underlying source charge $\varrho(s)$, one may assume that $s \ll r$. Fullerton and Rinker~\cite{fullerton1976} outlined a treatment for the Uehling potential which may be generalized for any vacuum polarization potential of the form we have defined in Eq.~(\ref{eq:vp-point-potential}). One need only take $|r-s| = r-s$ in Eq.~(\ref{eq:vp-potentials-finite}), partition the exponential functions as presented in Eq.~(\ref{eq:chi-functions}), and perform a Taylor series expansion near $s \approx 0$ on the resulting entity. Identifying the integrals over source charge $s$ with radial charge moments, we have
\begin{eqnarray}\label{eq:vp-fr-potential}
\lefteqn{\widetilde{V}_{m,n} (r)\bigr|_{r \to \infty} \simeq - \bigl( \tfrac{\alpha}{\pi} \bigr)^{m} (Z \alpha)^{n} \tfrac{1}{\alpha r} \Bigl[ \chi_{m,n} \bigl( \tfrac{2r}{\lambdabar} ; 1\bigr)}\\ \notag
 && \hspace{0.5cm} + \tfrac{2 R^{2}}{3 \lambdabar^{2}} \; \chi_{m,n} \bigl( \tfrac{2r}{\lambdabar} ; -1 \bigr) + \tfrac{2 \langle r^{4} \rangle}{15 \lambdabar^{4}} \; \chi_{m,n} \bigl( \tfrac{2r}{\lambdabar} ; -3 \bigr) + \ldots \Bigr].
\end{eqnarray}
The first term in this expansion returns the structureless point-charge result of Eq.~(\ref{eq:vp-point-potential}). This result is expected from Gauss's law of electrostatics, since the long-range behaviour is insensitive to the detailed structure of the source charge. The additional terms are suppressed not only by the functional form of the auxiliary functions for arguments $k < 0$, but also by unitless parameters such as $R/\lambdabar \approx 0.02$ given a typical nuclear charge radius $R \approx 6$ fm. The expansion relies on the availability of odd-valued functions as in Eq.~(\ref{eq:chi-functions}), which may be generated from the functions $w_{m,n}(t)$.

\subsection{Vacuum charge densities}\label{subsec:charges}

The renormalized vacuum polarization potentials are local functions of three-dimensional space, and may be directly associated with vacuum polarization charge densities through the Poisson equation:
\begin{equation}\label{eq:poisson-differential}
\nabla^{2} \widetilde{V}_{m,n}(r) = \tfrac{1}{r} \tfrac{\partial^{2}}{\partial r^{2}} \Bigl( r \widetilde{V}_{m,n}(r) \Bigr) = - 4 \pi \widetilde{\varrho}_{m,n}(r).
\end{equation}
A direct mapping may be established between a source charge density, $\varrho(r)$, and the induced vacuum density $\widetilde{\varrho}_{m,n}(r)$ generated by fluctuations of a charged virtual field at a particular diagrammatic order. Careful application of Leibniz' theorem leads to the result
\begin{widetext}
\begin{equation}\label{eq:vacpol-dens}
4 \pi r^{2} \widetilde{\varrho}_{m,n}(r)
  = \bigl( \tfrac{\alpha}{\pi} \bigr)^{m} (Z \alpha)^{n} \tfrac{1}{\alpha} \int_{\beta}^{\infty} \tfrac{1}{t} w_{m,n} (t) \biggl\{ - 4 \pi r^{2} \varrho(r) + \tfrac{4 \pi r t}{\lambdabar} \int_{0}^{\infty} s \varrho(s) \biggl[ \exp \Bigl( - \tfrac{2|r-s|t}{\lambdabar} \Bigr) - \exp \Bigl( - \tfrac{2(r+s)t}{\lambdabar} \Bigr) \biggr] \; \mathrm{d}s \biggr\} \; \mathrm{d}t.
\end{equation}
\end{widetext}
We have found that the direct application of Eq.~(\ref{eq:vacpol-dens}) is of limited utility. Instead, we utilize the form factor, $G(q)$, for the source charge in the relation
\begin{eqnarray}\label{eq:vp-dens-conv}
4 \pi r^{2} \widetilde{\varrho}_{m,n}(r)
 &=& 4 \pi r^{2} \mathcal{F}^{-1} \left[ G(q) \Pi_{m,n}(\lambdabar q) \right] \\ \notag
 &=& \tfrac{2 r^{2}}{\pi} \int_{0}^{\infty} q^{2} G(q) \Pi_{m,n} (\lambdabar q) j_{0}(rq) \; \mathrm{d}q.
\end{eqnarray}

The long-range expansion of the potential in Eq.~(\ref{eq:vp-fr-potential}) may be transformed into relations for the associated charge densities, by applying the differential form of the Poisson equation given in Eq.~(\ref{eq:poisson-differential}):
\begin{eqnarray*}
\lefteqn{4 \pi r^{2} \widetilde{\varrho}_{m,n}(r)\bigr|_{r \to \infty} \simeq (\tfrac{\alpha}{\pi})^{m} (Z \alpha)^{n} \tfrac{4r}{\alpha \lambdabar^{2}} \Bigl[ \chi_{m,n} \bigl( \tfrac{2r}{\lambdabar} ; -1\bigr)}\\
 && \hspace{0.5cm} + \tfrac{2 R^{2}}{3 \lambdabar^{2}} \; \chi_{m,n} \bigl( \tfrac{2r}{\lambdabar} ; -3 \bigr) + \tfrac{2 \langle r^{4} \rangle}{15 \lambdabar^{4}} \; \chi_{m,n} \bigl( \tfrac{2r}{\lambdabar} ; -5 \bigr) + \ldots \Bigr]
\end{eqnarray*}

The radial moments of the components of the vacuum polarization charge density, $\langle \widetilde{r}^{k} \rangle_{m,n} $, are given by 
\begin{equation}\label{eq:vp-chargemoment-integral}
\langle \widetilde{r}^{k} \rangle_{m,n} = \int_{0}^{\infty} 4 \pi r^{2+k} \widetilde{\varrho}_{m,n}(r) \; \mathrm{d}r.
\end{equation}
Rather than examining induced radial charge moments on a case-by-case basis, we have adapted methods that are used to determine nuclear charge moments. We begin with a power series expansion of the integrand in Eq.~(\ref{eq:transform-pair}),
\begin{equation}\label{eq:vp-cauchy-expand}
\tfrac{1}{t} \tfrac{u^{2}}{u^{2}+4t^{2}}\bigl|_{u \to 0} \simeq \tfrac{u^{2}}{4t^{3}} - \tfrac{u^{4}}{16 t^{5}} + \tfrac{u^{6}}{64 t^{7}} - \cdots.
\end{equation}
This facilitates the Taylor series expansion of the momentum space quantity $W_{m,n}(u)$ in the form
\begin{eqnarray}\notag
W_{m,n}(u)\bigl|_{u \to 0} 
& \simeq & - \tfrac{u^{2}}{4} \chi_{m,n} (0;3) + \tfrac{u^{4}}{16} \chi_{m,n} (0;5) \\ \label{eq:W-to-chi}
& & \hspace{0.5cm} - \tfrac{u^{6}}{64} \chi_{m,n} (0;7) + \cdots.
\end{eqnarray}
The properties of form factors as $q \to 0$ and the absence of a term of order unity in this expansion demonstrates explicitly that no net charge can be embedded in any of these potentials. This must be true for \textit{any} vacuum polarization potential which follows the general form of Eq.~(\ref{eq:vp-point-potential}), demonstrating that Ward's identity is satisfied for each of the diagrams that we examine here. \\

For $q\to 0$, a spherical Bessel function possesses the series expansion
\begin{eqnarray}\label{eq:inducedformfactor-powerseries}
\widetilde{G}_{m,n}(q)
 &=& \int_{0}^{\infty} 4 \pi r^{2} \widetilde{\varrho}_{m,n}(r) j_{0}(rq) \; \mathrm{d}r \\ \notag
 &=& \langle \widetilde{r}^{0} \rangle_{m,n} - \tfrac{1}{6} q^{2} \langle \widetilde{r}^{2} \rangle_{m,n} + \tfrac{1}{120} q^{4} \langle \widetilde{r}^{4} \rangle_{m,n}\\ \notag
 & &  \hspace{0.5cm} - \tfrac{1}{5040} q^{6} \langle \widetilde{r}^{6} \rangle_{m,n} + \cdots.
\end{eqnarray}
Combining this result with the expansion of $W(u)$ in Eq.~(\ref{eq:vp-cauchy-expand}) leads to the even-valued charge moments
\begin{eqnarray}\label{eq:vp-moments-0}
\langle \widetilde{r}^{0} \rangle_{m,n} &=& 0 \\ \notag
\langle \widetilde{r}^{2} \rangle_{m,n} &=& \tfrac{3 \lambdabar^{2}}{2} \bigl( \tfrac{\alpha}{\pi} \bigr)^{m} \alpha^{n-1} \chi_{m,n} \bigl(0 ; 3 \bigr) \\ \notag
\langle \widetilde{r}^{4} \rangle_{m,n} &=& \tfrac{15 \lambdabar^{4}}{2} \bigl( \tfrac{\alpha}{\pi} \bigr)^{m} \alpha^{n-1} \bigl[ \chi_{m,n} \bigl(0 ; 5\bigr) + \tfrac{2R^{2}}{3\lambdabar^{2}} \chi_{m,n}\bigl(0; 3 \bigr) \bigr]\\ \notag
\langle \widetilde{r}^{6} \rangle_{m,n} &=& \tfrac{315 \lambdabar^{6}}{4} \bigl( \tfrac{\alpha}{\pi} \bigr)^{m} \alpha^{n-1} \bigl[ \chi_{m,n} \bigl(0 ; 7\bigr)\\ \notag
 & & \hspace{1.5cm} + \tfrac{2R^{2}}{3\lambdabar^{2}} \chi_{m,n} \bigl(0 ; 5\bigr) + \tfrac{2\langle r^{4} \rangle}{15\lambdabar^{4}} \chi_{m,n} \bigl(0 ; 3\bigr) \bigr].
\end{eqnarray}
A few specific values of the auxiliary functions $\chi_{m,n}(0;k)$ may be used to determine global characteristics of the vacuum polarization charge densities which influence electronic structure. Should the function $W(u)$ be available and its expansion about $u=0$ known, then the coefficients of that expansion may be used directly rather than special values of the $\chi_{m,n}(0;k)$ functions.\\

For a structureless point source charge, $R=0$ and $\langle r^{4} \rangle = 0$, so that Eq.~(\ref{eq:vp-moments-0}) reveals that the vacuum polarization charge moments are finite and that these terms dominate if a finite nuclear model is employed. The zero- and second-order charge moments of vacuum polarization densities are insensitive to the details of the source charge. This does not suggest, however, that finite nuclear structure should be ignored entirely. For a point-like source, the generalized polarization charge density is of the form
\begin{eqnarray}\label{eq:induced-charge-structureless}
\lefteqn{4 \pi r^{2} \widetilde{\varrho}^{(0)}_{m,n}(r)} \\ \notag
 & = (\tfrac{\alpha}{\pi})^{m} (Z \alpha)^{n} \tfrac{1}{\alpha}
\Bigl[ \tfrac{4r}{\lambdabar^{2}} \chi_{m,n} \bigl( \tfrac{2r}{\lambdabar} ; -1 \bigr) - \chi_{m,n}(0;1) \delta (r) \Bigr].
\end{eqnarray}
This involves two charge distributions; one finite density which decays exponentially from the location of the source, and another which precisely counter-balances this finite charge density, confined as a point response function at the origin. By making the source charge finite and smooth, this abrupt behaviour is replaced by an outer region with one sign in charge, perfectly counter-balanced by an inner region with the other sign in charge.

\section{Implementation}\label{sec:implementation}

We have established that the green path in Fig.~\ref{fig:vp-connections} provides a convenient and readily generalizable approach to the calculation of vacuum polarization phenomena. This route requires only the existence of a nuclear form factor $G(q)$, which is readily derived from a given nuclear charge model, and a representation of the spectral function $\Pi_{m,n}(\lambdabar q)$ for the virtual field and free-field diagram. We often find that the required spectral function not only exists in closed form, but may be evaluated with relative ease. Where we have not been able to determine a closed-form expression for the spectral function, it may be evaluated \textit{numerically} by means of Eq.~(\ref{eq:transform-pair}), provided that the associated function $w_{m,n}(t)$ is available.\\

The dominant electrostatic sources within molecular systems that polarize the vacuum are nuclear charges. It is the coupling of those nuclei to the virtual electron-positron fields that produce vacuum polarization potentials which impact electronic bound states within those systems. Here, we outline the construction of those potentials for cases where the underlying nucleus comprises a spherically symmetric and finite charge density $\varrho(r)$.

\subsection{Electron-positron field, order $\alpha(Z\alpha)$}\label{subsec:nuclei_e_aZa}

The influence of nuclei on the fermionic electron-positron field at leading order is known as the Uehling interaction. We have previously described our approach to calculating the corresponding potential, so here we simply summarize its main features.\\

The Uehling potential is generated using
\begin{equation}\label{eq:f_11_e}
f_{1,1} (t) = \tfrac{2}{3} \sqrt{1 - \tfrac{1}{t^{2}}} \left( 1 + \tfrac{1}{2t^{2}} \right).
\end{equation}
The auxiliary functions $\chi_{1,1}^{f}(x,k)$, defined by Eq.~(\ref{eq:chi-functions}), have absorbed an additional factor of $2/3$ compared to their conventional representations. Some useful special values of those functions are
\begin{equation}\label{eq:chi_n_0}
\chi_{1,1}^{f}(0;k) = \tfrac{\sqrt{\pi}}{2(k-1)} \tfrac{\Gamma ( (k+3)/2 )}{\Gamma ( (k+4)/2 )} \quad \text{if} \quad k > 1.
\end{equation}
A point-like nucleus $Z$ generates a structureless potential which diverges logarithmically at the origin, which can be seen by comparing Eq.~(\ref{eq:vp-point-potential}) with Eq.~(\ref{eq:chi_n_0}). In the long-range asymptotic region for which $r \gg \lambdabar_{e}$, we may approximate the potential by a modified exponential function,
\begin{equation}
\widetilde{V}^{(0,e)}(r) \bigr|_{r \to \infty} \simeq
 - \tfrac{Z\alpha}{\pi r} \sqrt{\tfrac{\pi}{2}} \bigl( \tfrac{2r}{\lambdabar_{e}} \bigr)^{-3/2} e^{-2r/\lambdabar_{e}} \end{equation}
which effectively cuts off the interaction.\\

The momentum space representation for this interaction was presented by Tsai~\cite{tsai1960}. In our notation this involves the function
\begin{equation}
F_{1,1}(u) = - \tfrac{5}{6} + \tfrac{2}{u^{2}} + \tfrac{1}{u^{3}} (u^{2}-2) \sqrt{u^{2}+4} \; \text{arcsinh}  \bigl(\tfrac{u}{2} \bigr).
\end{equation}
In the ultraviolet (high-$u$) limit, Maximon~\cite{maximon2000} showed that
\begin{equation}
F_{1,1} (u) \approx - \tfrac{5}{6} + \log (u) + \tfrac{3}{u^{2}} + \tfrac{1}{u^{4}} \left( - \tfrac{3}{2} + 6 \log (u) \right).
\end{equation}

In the limit $u \to 0$, this function can be expanded as power series:
\begin{equation}
F_{1,1} (u) \approx \tfrac{1}{10} u^{2} - \tfrac{3}{280} u^{4} + \tfrac{1}{630} u^{6} - \tfrac{1}{3696} u^{8}. 
\end{equation}
By comparing this expansion with the induced form factor in Eq.~(\ref{eq:inducedformfactor-powerseries}) for the electron-positron field, we obtain the vacuum polarization radial charge moments
\begin{subequations}\label{eq:f11-even-moments}
\begin{eqnarray}\label{eq:f11-r0}
\langle \widetilde{r}^{0} \rangle_{1,1}^{e}
 &=& 0 \\ \label{eq:f11-r2}
\langle \widetilde{r}^{2} \rangle_{1,1}^{e}
 &=& \tfrac{2\alpha\lambdabar^{2}_{e}}{5 \pi} \\ \label{eq:f11-r4}
\langle \widetilde{r}^{4} \rangle_{1,1}^{e}
 &=& \tfrac{6 \alpha \lambdabar^{4}_{e}}{7 \pi} \bigl( 1 + \tfrac{14R^{2}}{9 \lambdabar^{2}_{e}} \bigr) \\ \label{eq:f11-r6}
\langle \widetilde{r}^{6} \rangle_{1,1}^{e}
 &=& \tfrac{16 \alpha \lambdabar^{6}_{e}}{3 \pi} \bigl( 1 + \tfrac{9R^{2}}{8 \lambdabar^{2}_{e}} + \tfrac{21\langle r^{4} \rangle}{40 \lambdabar^{4}_{e}} \bigr).
\end{eqnarray}
\end{subequations}
This demonstrates explicitly that the induced charge density carries no net charge, and that its root-mean-square radius is not sensitive to the details of the source charge. Similarly, the higher induced charge moments are only weakly dependent on the \textit{lower} charge moments of the sources.
\begin{figure}[!h]
\begin{center}
\includegraphics[width=8.5cm]{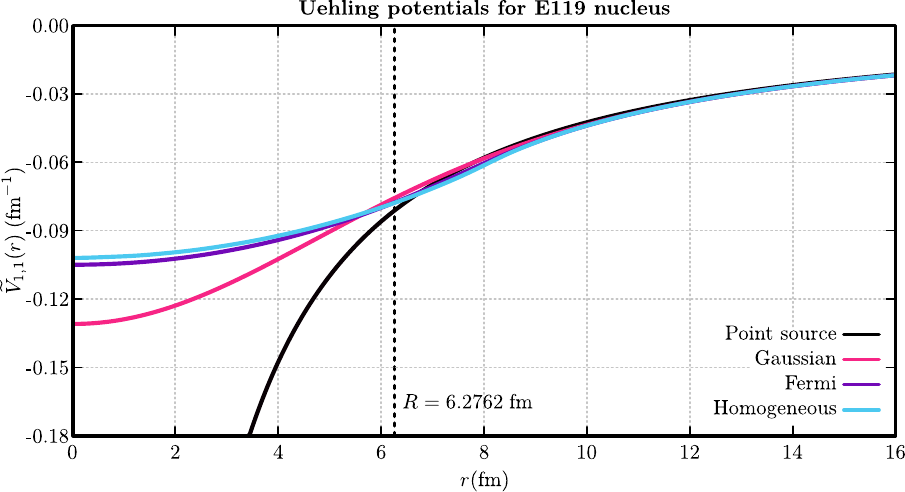}
\includegraphics[width=8.5cm]{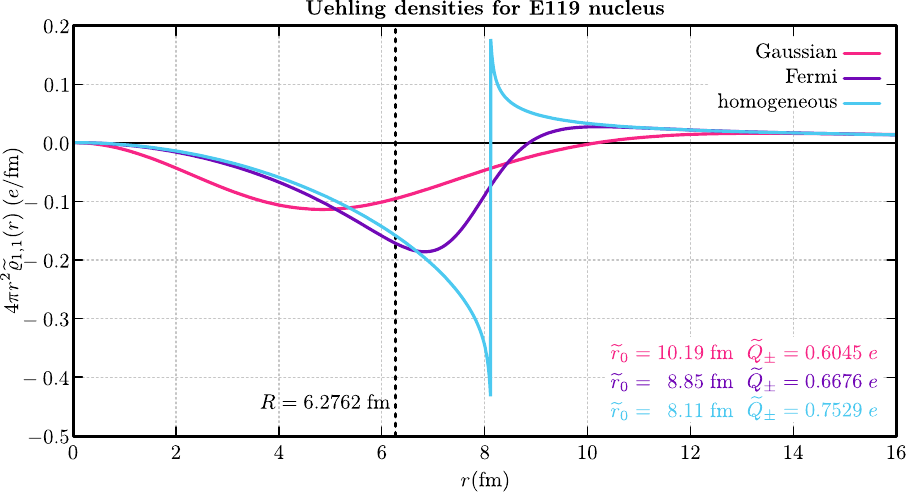}
\caption{The Uehling potential and associated charge density for an E119 nucleus with radius $R = 6.2762 \;$fm, for a series of nuclear charge models.}
\label{fig:vp-E119-potential-f11}
\end{center}
\end{figure}
\vspace{-0.6cm}
\subsection{Electron-positron field, order $\alpha(Z\alpha)^{3}$}\label{subsec:nuclei_e_aZa3}

The order $\alpha (Z \alpha)^{3}$ contribution is known as the Wichmann-Kroll interaction~\cite{wichmann1956}. It becomes significant for heavy elements and strong Coulomb fields, for which $Z\alpha$ is not a small parameter. The relevant energy-like function required to generate a Wichmann-Kroll potential is defined on the range $0 \leq t < \infty$, which means that $\beta = 0$ in Eq.~(\ref{eq:chi-functions}), and
\begin{equation*}
f_{1,3} (t) 
 = \tfrac{\pi^{2}}{12 t^{2}} \sqrt{1 - \tfrac{1}{t^{2}}} \Theta(t-1)
 - \tfrac{1}{t^{2}} \int_{0}^{t} \sqrt{1-\tfrac{x^{2}}{t^{2}}} X(x) \; \mathrm{d}x,
\end{equation*}
\begin{widetext}
\begin{subequations}\label{eq:Xboundaries}
\begin{eqnarray} \notag
X_{<}(x)
 &=& -2x \mathrm{Li}_{2}(x^{2}) - x \log^{2} (1-x^{2}) + \tfrac{1-x^{2}}{x^{2}} \log (1- x^{2}) \log \bigl( \tfrac{1+x}{1-x} \bigr) \\
 & & + \tfrac{1-x^{2}}{4x} \log^{2} \bigl( \tfrac{1+x}{1-x} \bigr) + \tfrac{2 - x^{2}}{x(1-x^{2})} \log (1-x^{2}) + \tfrac{3-2x^{2}}{1-x^{2}} \log \bigl(  \tfrac{1+x}{1-x} \bigr) - 3x \\
 X_{>}(x) \notag
 &=& x^{-2} \mathrm{Li}_{2}(x^{-2}) - \tfrac{3x^{2}+1}{2x} \left[ \mathrm{Li}_{2}(x^{-1}) - \mathrm{Li}_{2}(- x^{-1}) \right] - \tfrac{2x^{2}-1}{2x^{2}} \bigl[ \log^{2} (1-x^{-2}) + \log^{2} \bigl( \tfrac{x+1}{x-1} \bigr) \bigr] \\ \notag
 & & - (2x-1) \log (1-x^{-2}) \log \bigl( \tfrac{x+1}{x-1} \bigr) +  \tfrac{3x^{2}+1}{4x} \log^{2} \bigl( \tfrac{x+1}{x-1} \bigr) - 2 \log (x) \log (1-x^{-2}) - \tfrac{3x^{2}+1}{2x} \log(x) \log \bigl( \tfrac{x+1}{x-1} \bigr) \\
 & & + \bigl( 5 - \tfrac{x(3x^{2}-2)}{x^{2}-1} \bigr) \log (1-x^{-2}) 
     + \bigl( \tfrac{3x^{2}+2}{x} - \tfrac{3x^{2}-2}{x^{2}-1} \bigr) \log  \bigl( \tfrac{x+1}{x-1} \bigr) + 3 \log(x) - 3.
\end{eqnarray}
\end{subequations}
\end{widetext}
This function involves the Heaviside step function, $\Theta (t-1)$, as well as a dimensionless quantity $X(x)$ which is most conveniently separated at $x=1$, with functions $X_{<}(x)$ and $X_{>}(x)$ on either side of the boundary as defined above in Eq.~(\ref{eq:Xboundaries}). Here Li$_{2}(x)$ is the dilogarithm function, defined by
\begin{equation*} 
\mathrm{Li}_{2}(x) = - \int_{0}^{x} \tfrac{\log(1-z)}{z} \; \mathrm{d}z = \sum_{k=1} ^{\infty} \tfrac{x^{k}}{k^{2}}, \qquad -1 \leq x \leq 1.
\end{equation*}
The integration over the variable $x$ that defines $f_{1,3}(t)$ is interpreted as a principal value integral. The behaviour of the integrand in the neighbourhood of $x=1$ was analysed using Mathematica, and the divergent terms eliminated, since they make no contribution to the principal value integral. The remaining part of the integrand in the neighbourhood of $x=1$ is smooth, and can be integrated numerically over the interval $1-\delta < x < 1+\delta$, for sufficiently small $\delta$.\\

Defining $a(r) = \log r+ \gamma$ where $\gamma \approx 0.5772$ is the Euler constant, Blomqvist \cite{blomqvist1972} derived a power series expansion up to $\mathcal{O}(r^{3})$ for this potential in the vicinity of the origin:
\begin{eqnarray}\label{eq:vp-wkr-near0}
\lefteqn{\widetilde{V}_{1,3}^{(0)}(r) \bigr|_{r \to 0} \simeq \tfrac{\alpha (Z \alpha)^{3}}{\pi r} \Bigl\{
     \bigl[ \tfrac{\pi^{2}}{6} - \tfrac{7}{9} - \tfrac{2}{3} \zeta(3) \bigr] }\\ \notag
 & & - \bigl[ \tfrac{\pi^{3}}{4} -  2 \pi \zeta(3) \bigr] r
     + \bigl[ \tfrac{\pi^{4}}{16} + \tfrac{\pi^{2}}{6} - 6\zeta(3)\bigr] r^{2} 
     + \tfrac{2 \pi}{9} a(r) r^{3} \\ \notag
 & & - \bigl[ \tfrac{31 \pi}{27} - \tfrac{4 \pi}{9} \log 2 - \tfrac{2 \pi}{3} \zeta(3) \bigr] r^{3}
     + \tfrac{1}{12} a(r)^{2} r^{4} \\ \notag
 & & + \bigl[ \tfrac{5 \pi^{2}}{54} - \tfrac{19}{36} \bigr] a(r) r^{4}
     - \bigl[ \tfrac{109 \pi^{2}}{432} + \tfrac{859}{864} - \tfrac{13}{18} \zeta(3) \bigr] r^{4}
  \Bigr\}.
\end{eqnarray}
This behaviour indicates that a finite point charge of $Q_{-} = -[\tfrac{\pi^{2}}{6} - \tfrac{7}{9} - \tfrac{2}{3} \zeta(3)](Z \alpha)^{3} \alpha / \pi \approx -0.0658 (Z \alpha)^{3} \alpha / \pi$ resides precisely on the origin, which was identified in Eq. (61) of Wichmann and Kroll~\cite{wichmann1956}. It is precisely counter-balanced by an extended positive charge density. The finiteness of this result provides a special value of the auxiliary function
\begin{equation}
\chi^{f}_{1,3}(0;1) = \tfrac{2}{3} \zeta(3)  + \tfrac{7}{9} - \tfrac{\pi^{2}}{6}.
\end{equation}

At large radial distances from the point source charge, the structureless Wichmann-Kroll potential behaves like 
\begin{eqnarray}\label{eq:vp-wk-longdistance}
\widetilde{V}^{(0)}_{1,3}(r) 
 &\approx & \tfrac{(Z \alpha)^{3} \alpha}{\pi} \tfrac{1}{r^{5}}
 \bigl( \tfrac{2}{225} + \tfrac{59}{1323} \tfrac{1}{r^{2}} + \tfrac{659}{1575} \tfrac{1}{r^{4}} \\ \notag
 && \hspace{0.5cm} + \tfrac{659}{1575} \tfrac{1}{r^{6}} + \tfrac{83912}{12705} \tfrac{1}{r^{8}} + \tfrac{217824448}{1366365 } \tfrac{1}{r^{10}} + \cdots \bigr),
\end{eqnarray}
for which the associated charge density is
\begin{eqnarray}\label{eq:vp-wkr-densinf}
\lefteqn{4 \pi r^{2} \widetilde{\varrho}^{(0)}_{1,3} (r) 
 \approx -\tfrac{(Z \alpha)^{3} \alpha}{\pi} \tfrac{1}{r^{5}} \bigl( \tfrac{8}{45} + \tfrac{118}{63} \tfrac{1}{r^{2}} + \tfrac{5272}{175} \tfrac{1}{r^{4}}} \\ \notag
 & \hspace{1.0cm} + \tfrac{167824}{231} \tfrac{1}{r^{6}} + \tfrac{871297792}{35035} \tfrac{1}{r^{8}} + \tfrac{1484876352}{143} \tfrac{1}{r^{10}} + \cdots  \bigr).
\end{eqnarray}
Since asymptotic behaviour of the Wichmann-Kroll charge density depends on $1/r^{5}$, this imposes an apparent restriction on the convergence of the charge moments, because on this basis they should diverge for $k\geq 4$. This behaviour can be traced to the general form of induced radial charge moments in Eq.~(\ref{eq:vp-moments-0}) and definition of the auxiliary functions in Eq.~(\ref{eq:chi-functions}): since the function $f_{1,3}(t)$ is defined on the range $t: [0,\infty)$ and its behaviour near $t=0$ is proportional to $t^{4}$, any value $k \geq 4$ will result in divergent special values $\chi^{f}_{1,3}(0;k)$.\\

The charge moments of the Wichmann-Kroll density may be identified directly, by means of Eq.~(\ref{eq:vp-chargemoment-integral}) and an integration scheme that imposes a hard radial cut-off near the onset of the asymptotic power law behaviour at $r=7.5 \lambdabar_{e}$. This results in the estimated relations
\begin{subequations}\label{eq:f13-even-moments}
	\begin{eqnarray}\label{eq:f13-r0}
		\langle \widetilde{r}^{0} \rangle_{1,3}^{e}
		&=& 0 \\ \label{eq:f13-r2}
		\langle \widetilde{r}^{2} \rangle_{1,3}^{e}
		&=& -\tfrac{9\alpha^{3}\lambdabar^{2}_{e}}{106 \pi} \\ \label{eq:f13-r4}
		\langle \widetilde{r}^{4} \rangle_{1,3}^{e}
		&=& -\tfrac{15 \alpha^{3} \lambdabar^{4}_{e}}{4 \pi} \bigl( 0.0721 + \tfrac{2R^{2}}{53 \lambdabar^{2}_{e}} \bigr) \\ \label{eq:f13-r6}
		\langle \widetilde{r}^{6} \rangle_{1,3}^{e}
		&=& -\tfrac{315\alpha^{3} \lambdabar^{6}_{e}}{4\pi} \bigl( 0.1128 + 0.0721 \tfrac{2R^{2}}{3\lambdabar^{2}_{e}} + \tfrac{2\langle r^{4} \rangle}{53\lambdabar^{4}_{e}} \bigr)  \ \
	\end{eqnarray}
\end{subequations}
In the absence of a spectral function for the Wichmann-Kroll interaction, we evaluated $F_{1,3}(u)$ from $f_{1,3}(t)$ by means of numerical quadrature on
\begin{equation}
F_{1,3}(u) = \int_{0}^{\infty} \tfrac{1}{t} \tfrac{u^{2}}{u^{2} + 4 t^{2}} f_{1,3}(t) \; \mathrm{d}t,
\end{equation}
the results being saved to an external file for later use in generating the appropriate potentials using Eq.~(\ref{eq:polarization-potential-transform}).\\

Wichmann and Kroll~\cite{wichmann1956} also discussed the properties of a potential $\widetilde{V}^{(0)}_{1,3+}(r)$ which accounts for all remaining terms in the $(Z \alpha)$ expansion at order $\alpha$. Calculations of this potential were made by Johnson and Soff \cite{johnson1985} with analytical details discussed by Fainshtein, Manakov and Nekipelov \cite{fainshtein1991}, demonstrating that $\widetilde{V}^{(0)}_{1,3+}(r)$ gives rise to corrections on the order of only a few percent of the conventional Wichmann-Kroll interaction in hydrogen-like atomic systems.
\begin{figure}[!h]
    \begin{center}
    \includegraphics[width=8.5cm]{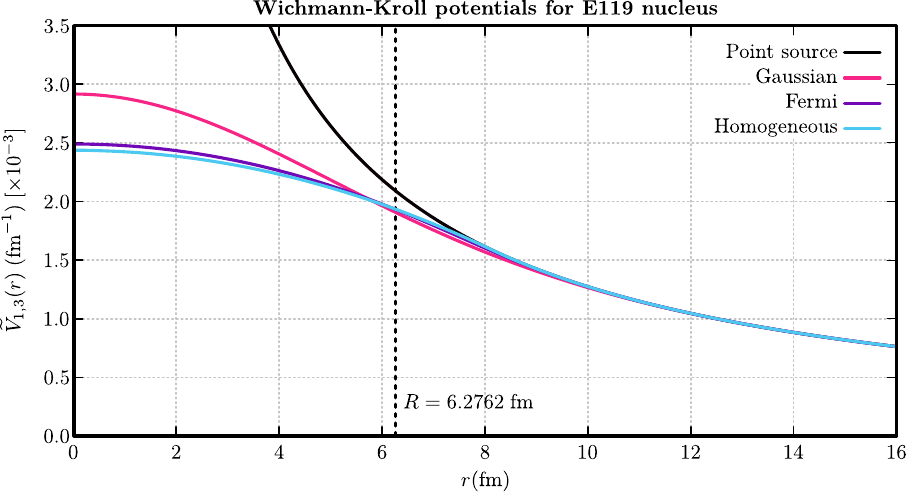}
    \includegraphics[width=8.5cm]{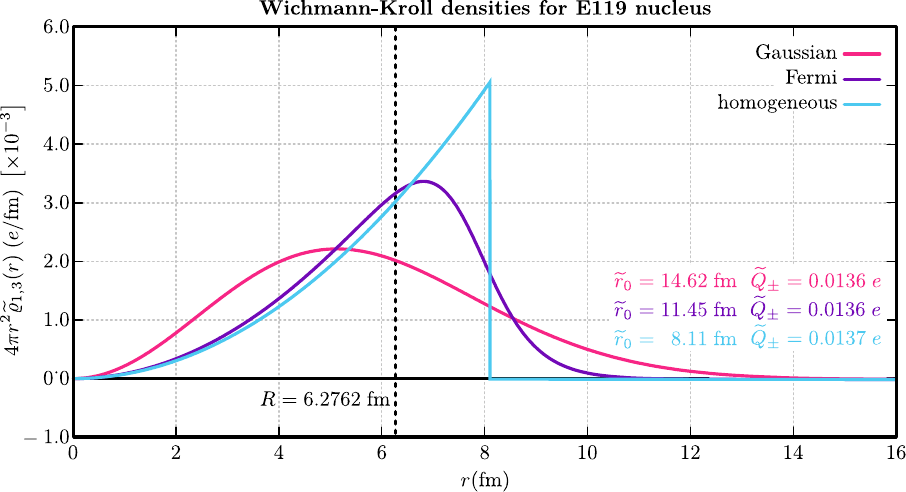}
    \caption{The Wichmann-Kroll potential and associated charge density for an E119 nucleus with radius $R = 6.2762 \;$fm, for a series of nuclear charge models.}
    \label{fig:vp-E119-potential-f13}
    \end{center}
\end{figure}
\vspace{-0.6cm}
    
\subsection{Electron-positron field, order $\alpha^{2}(Z\alpha)$}\label{subsec:nuclei_e_a2Za}

For the bound-state vacuum polarization diagrams that involve the exchange of two virtual photons (a subset of all the order-$\alpha^{2}$ diagrams), the free-field expansion of those diagrams in powers of $(Z \alpha)^{2n-1}$ involves a number of terms. One of those terms is a ladder diagram involving two distinct loops, but since that diagram is understood to arise naturally from inclusion of the one-loop diagram, it is partitioned from the K\"{a}ll\'{e}n and Sabry~\cite{kallen1955} definition of the $\alpha^{2}(Z\alpha)$ interaction. The other terms may be handled with the function
\begin{eqnarray}
f_{2,1}(t) 
 &=& -\bigl( \tfrac{13}{54} + \tfrac{7}{108 t^{2}} + \tfrac{2}{9t^{4}} \bigr) \sqrt{1 - \tfrac{1}{t^{2}}} \\ \notag
 & &+ \left( \tfrac{44}{9} - \tfrac{2}{3t^{2}} - \tfrac{5}{4t^{4}} - \tfrac{2}{9t^{6}} \right) \log \bigl( t + \sqrt{t^{2}-1} \bigr) \\ \notag
 & & - \bigl( \tfrac{4}{3} + \tfrac{2}{3t^{2}} \bigr) \sqrt{1 - \tfrac{1}{t^{2}}} \log \bigl( 8t (t^{2}-1) \bigr) \\ \notag
 & & + \bigl( \tfrac{8}{3} - \tfrac{2}{3t^{4}} \bigr) \int_{t}^{\infty} \tfrac{3x^{2}-1}{x(x^{2}-1)} \log \bigl( x + \sqrt{x^{2}-1} \bigr) \\ \notag
 & & - \tfrac{1}{\sqrt{x^{2}-1}} \log \bigl( 8x (x^{2}-1) \bigr) \; \mathrm{d}x.
\end{eqnarray}
Near the origin and again taking $a(r) = \log(r)+\gamma$, a point-like source charge generates a K\"{a}ll\'{e}n-Sabry vacuum polarization potential
\begin{equation}\label{eq:vp-ksb-near0}
\widetilde{V}_{2,1}^{(0)}(r) \approx -\tfrac{(Z \alpha) \alpha^{2}}{\pi^{2}} \tfrac{1}{r} v(r),
\end{equation}
where
\begin{eqnarray*}
    v(r)
     &=& \tfrac{\pi^{2}}{27} + \tfrac{65}{648} - (\tfrac{13 \pi^{2}}{9} + \tfrac{32 \pi}{9} \log 2 - \tfrac{766 \pi}{135}) r \\ \notag
     && + (\tfrac{65}{18} - \tfrac{5}{3} a(r)) r^{2} + (\tfrac{80 \pi}{81} - \tfrac{14 \pi^{2}}{27}) r^{3} \\ \notag
     && + (\tfrac{5}{18} a(r)^{2} + \tfrac{323}{216} a(r)) r^{4} - \bigl( \tfrac{1}{6} \zeta(3) - \tfrac{5 \pi^{2}}{216} - \tfrac{6509}{2592}\bigr) r^{4}.
\end{eqnarray*}
Like the Wichmann-Kroll interaction, a point charge and a counterbalancing distribution are induced in the vacuum surrounding a point source charge through the K\"{a}ll\'{e}n-Sabry interaction.\\

K\"{a}ll\'{e}n and Sabry~\cite{kallen1955} also provided the appropriate spectral function for these diagrams:
\begin{eqnarray*}
F_{2,1}(u)
    &=& \bigl( \tfrac{13}{324} - \tfrac{11}{216} \delta^{2} + \tfrac{1}{9} \delta^{4} \bigr)\\
    & & + \bigl( -\tfrac{19}{72} + \tfrac{55}{216} \delta^{2} - \tfrac{1}{9} \delta^{4} \bigr) \delta \log \bigl( \tfrac{1+\delta}{|1-\delta|} \bigr)\\
    & & + \bigl( -\tfrac{11}{32} - \tfrac{23}{48} \delta^{2} + \tfrac{23}{96} \delta^{4} - \tfrac{1}{36} \delta^{6} \bigr) \log^{2} \bigl( \tfrac{1+\delta}{|1-\delta|} \bigr) \\
    & &  - \delta (1 - \tfrac{1}{3}\delta^{2}) \Bigl[ \Phi \bigl( \tfrac{1 - \delta}{1 + \delta} \bigr) + 2 \Phi \bigl(- \tfrac{1 - \delta}{1 + \delta} \bigr) + \tfrac{\pi^{2}}{4}\\
    & & - \tfrac{3}{4} \log^{2} \bigl( \tfrac{1+\delta}{|1-\delta|} \bigr) + \tfrac{1}{2} \log \bigl( \tfrac{1+\delta}{|1-\delta|} \bigr) \log \bigl( \tfrac{64 \delta^{2}}{1-\delta^{2}|^{3}} \bigr) \Bigr]\\
    & &  - (1 + \tfrac{2}{3} \delta^{2} - \tfrac{1}{3}\delta^{4})  \int_{-1}^{+1} \log \bigl| 1 - \tfrac{x^{2}}{\delta^{2}} \bigr| I(x) \; \mathrm{d}x
\end{eqnarray*}
where
\begin{equation*}
I(x) = \bigl[ \tfrac{1}{x} \log(|1+x|) + \tfrac{3}{2} \tfrac{1}{1+x} \log \bigl( \tfrac{1-x}{2} \bigr)
- \tfrac{1}{1+x} \log (|x|) \bigr].
\end{equation*}
Defining the variable $\delta(u) = \sqrt{1+4/u^{2}}$ and an auxiliary function
\begin{equation}
\Phi (x) = \int_{1}^{x} \tfrac{1}{t} \log \bigl(|1+t|\bigr) \; \mathrm{d}t,
\end{equation}
a Taylor expansion of $F_{2,1} (u)$ in the limit $u \to 0$ yields
\begin{equation}
F_{2,1} (u) \approx \tfrac{41}{162} u^{2} - \tfrac{401}{10800} u^{4} + \tfrac{5919}{7938000} u^{6} - \tfrac{25993}{16329600} u^{8}.
\end{equation}
This corresponds precisely to each of the dominant terms in our calculated charge moments for the associated charge density $\widetilde{\varrho}_{2,1}(r)$,
\begin{subequations}\label{eq:f21-even-moments}
\begin{eqnarray}\label{eq:f21-r0}
\langle \widetilde{r}^{0} \rangle_{2,1}^{e}
 &=& 0 \\ \label{eq:f21-r2}
\langle \widetilde{r}^{2} \rangle_{2,1}^{e}
 &=& \tfrac{41\alpha^{2}\lambdabar^{2}_{e}}{27 \pi^{2}} \\ \label{eq:f21-r4}
\langle \widetilde{r}^{4} \rangle_{2,1}^{e}
 &=& \tfrac{401 \alpha^{2} \lambdabar^{4}_{e}}{90 \pi^{2}} \bigl( 1 + \tfrac{4100R^{2}}{3609 \lambdabar^{2}_{e}} \bigr) \\ \label{eq:f21-r6}
\langle \widetilde{r}^{6} \rangle_{2,1}^{e}
 &=& \tfrac{54919\alpha^{2} \lambdabar^{6}_{e}}{1575\pi^{2}} \bigl( 1 + \tfrac{98245 R^{2}}{109388\lambdabar^{2}_{e}} + \tfrac{50225\langle r^{4} \rangle}{164757\lambdabar^{4}_{e}} \bigr).
\end{eqnarray}
\end{subequations}

\begin{figure}[!t]
\begin{center}
\includegraphics[width=8.5cm]{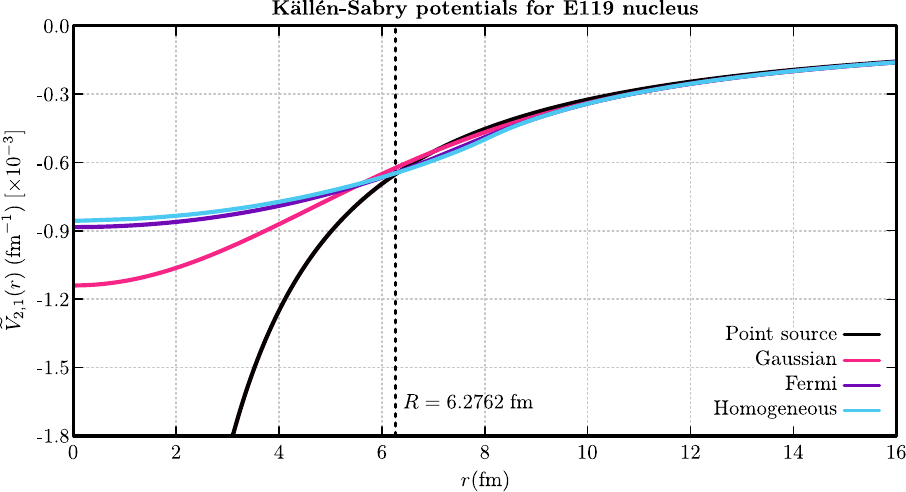}
\includegraphics[width=8.5cm]{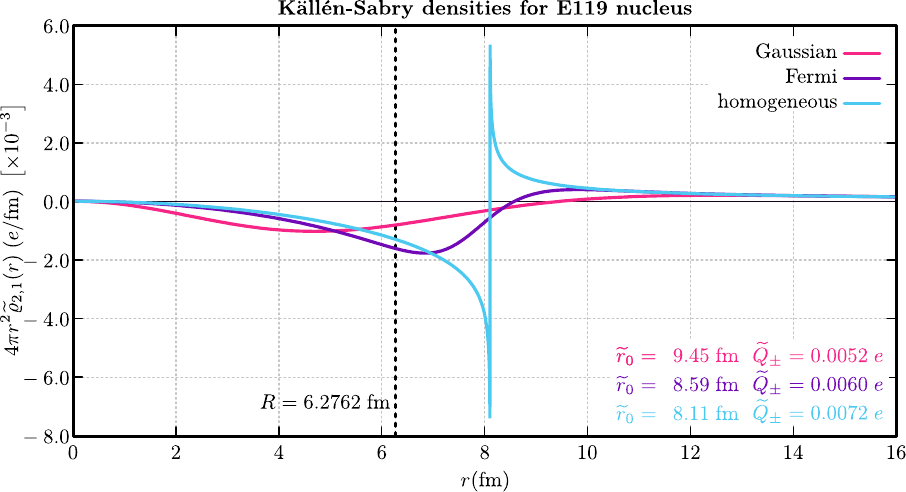}
\caption{The K\'{a}llen-Sabry potential and associated charge density for an E119 nucleus with radius $R = 6.2762 \;$fm, for a series of nuclear charge models.}
\label{fig:vp-E119-potential-f21}
\end{center}
\end{figure}

\subsection{Muon field, order $\alpha(Z\alpha)$}\label{subsec:nuclei_muon}

The Uehling potential that arises from the coupling between a nucleus and the virtual muon field may be obtained by modifying the Compton wavelength so that $\lambdabar_{\mu} = \lambdabar_{e}/(m_{\mu}/m_{e}) \approx \lambdabar_{e}/207$. This restricts the natural length range of any vacuum polarization effects to less than 2~fm, which lies well within the volume of any realistic nucleus.\\

The Uehling potential, induced charge density, radial charge moments and polarization form factor for this interaction at order $\alpha(Z\alpha)$ can be generated with the relations outlined in Eqs.~(\ref{eq:f_11_e})-(\ref{eq:f11-even-moments}). Of particular note is the spectral function for this interaction, which involves values $u=\lambdabar_{\mu}q$ that are particularly weighted towards the small-$u$ domain. As a result, radial charge moments of the induced density are much more sensitive to the details of the underlying source charge:
\begin{subequations}\label{eq:u11-even-moments}
\begin{eqnarray}\label{eq:u11-r0}
\langle \widetilde{r}^{0} \rangle_{1,1}^{\mu}
 &=& 0 \\ \label{eq:u11-r2}
\langle \widetilde{r}^{2} \rangle_{1,1}^{\mu}
 &=& \tfrac{2\alpha\lambdabar^{2}_{\mu}}{5 \pi} \\ \label{eq:u11-r4}
\langle \widetilde{r}^{4} \rangle_{1,1}^{\mu}
 &=& \tfrac{6 \alpha \lambdabar^{4}_{\mu}}{7 \pi} \bigl( 1 + \tfrac{14R^{2}}{9 \lambdabar^{2}_{\mu}} \bigr) \\ \label{eq:u11-r6}
\langle \widetilde{r}^{6} \rangle_{1,1}^{\mu}
 &=& \tfrac{16 \alpha \lambdabar^{6}_{\mu}}{3 \pi} \bigl( 1 + \tfrac{9R^{2}}{8 \lambdabar^{2}_{\mu}} + \tfrac{21\langle r^{4} \rangle}{40 \lambdabar^{4}_{\mu}} \bigr).
\end{eqnarray}
\end{subequations}

The hypothetical E119 nucleus, for instance, derives almost 95\% of the moment $\langle \widetilde{r}^{4} \rangle_{1,1}^{\mu}$ from the second term in the parentheses of Eq.~(\ref{eq:u11-r4}). Similarly, the Fullerton-Rinker expansion in Eq.~(\ref{eq:vp-fr-potential}) for a vacuum polarization potential at $r \gg \lambdabar_{\mu}$ loses its utility in this situation, since the ratio $R/\lambdabar > 1$ yields a divergent expansion. In practice this is not a problem, because the potential decays so quickly that it only need be evaluated over a short radial range, of the order of the nuclear radius $R$.

\subsection{Charged pion field, order $\alpha(Z\alpha)$}\label{subsec:nuclei_pion}

Vacuum polarization diagrams may be calculated for charge-carrying virtual fields other than the electron-positron field. In general, however, such interactions impart progressively weaker influences on bound-state electrons within an atom or molecule. We focus here on the leading-order Uehling diagram for the virtual pion field $\pi^{+}\pi^{-}$. These particles are bosons and require the use of a different effective QED potential than the fermionic electron-positron field. We follow the work of Tsai~\cite{tsai1960}, who derived the appropriate spectral function for boson fields,
\begin{equation}\label{eq:vp-ueh-spec-boson}
 B_{1,1}(u) = - \tfrac{2}{3} - \tfrac{2}{u^{2}} + \tfrac{1}{2u^{3}} \left( u^{2}+4 \right)^{3/2} \; \text{arcsinh} \bigl(\tfrac{u}{2} \bigr).
\end{equation}
This treatment involves the assumption that pions can be regarded as structureless particles, which is likely to be a good approximation in the calculation of small effects on bound-state electrons.\\

Maximon~\cite{maximon2000} showed that this function is characterized by a logarithmic divergence in the ultraviolet limit, 
\begin{eqnarray}
B_{1,1} (u) &\approx & - \tfrac{2}{3} + \tfrac{1}{2} \log (u)\\ \notag
 & & + \tfrac{1}{u^{2}} \bigl( - \tfrac{3}{2} + 3 \log (u) \bigr) + \tfrac{1}{u^{4}} \bigl( \tfrac{9}{4} + 3 \log (u) \bigr),
\end{eqnarray}
For $u \to 0$, we may expand the function in the power series
\begin{eqnarray}
B_{1,1} (u) \approx \tfrac{1}{40} u^{2} - \tfrac{1}{560} u^{4} + \tfrac{1}{5040} u^{6} - \tfrac{1}{3696} u^{8}.
\end{eqnarray}
The associated energy-like function for this diagram is simply
\begin{equation}
b_{1,1} (t) = \tfrac{1}{3} \bigl( 1 - \tfrac{1}{t^{2}} \bigr)^{3/2},
\end{equation}
with special values of the auxiliary function given by
\begin{equation}\label{eq:chi_special_b}
\chi_{1,1}^{b}(0;k) = \tfrac{\sqrt{\pi}}{4(k-1)} \tfrac{\Gamma ( (k+1)/2 )}{\Gamma ( (k+4)/2 )} \quad \text{if} \quad k > 1.
\end{equation}

The induced charge density produced by the redistribution of virtual $\pi^{+}\pi^{-}$ pairs involves a Compton wavelength $\lambdabar_{\pi} \approx \lambdabar_{e}/263$. The radial charge moments of that density are
\begin{subequations}\label{eq:b11-even-moments}
\begin{eqnarray}\label{eq:b11-r0}
\langle \widetilde{r}^{0} \rangle_{1,1}^{\pi}
 &=& 0 \\ \label{eq:b11-r2}
\langle \widetilde{r}^{2} \rangle_{1,1}^{\pi}
 &=& \tfrac{\alpha\lambdabar^{2}_{\pi}}{10 \pi} \\ \label{eq:b11-r4}
\langle \widetilde{r}^{4} \rangle_{1,1}^{\pi}
 &=& \tfrac{\alpha \lambdabar^{4}_{\pi}}{7 \pi} \bigl( 1 + \tfrac{7R^{2}}{3 \lambdabar^{2}_{\pi}} \bigr) \\ \label{eq:b11-r6}
\langle \widetilde{r}^{6} \rangle_{1,1}^{\pi}
 &=& \tfrac{2 \alpha \lambdabar^{6}_{\pi}}{3 \pi} \bigl( 1 + \tfrac{3R^{2}}{2 \lambdabar^{2}_{\pi}} + \tfrac{21\langle r^{4} \rangle}{20 \lambdabar^{4}_{\pi}} \bigr).
\end{eqnarray}
\end{subequations}
Just as we found for the polarization of the electron-positron field, the net induced charge vanishes and the root-mean-squared radius is insensitive to the details of the source charge. Both of these characteristics are critical in treating molecular electronic structure problems. The higher radial charge moments incorporate features of the source charge, and because of the particularly short Compton wavelength $\lambdabar_{\pi} \approx 1.5$fm, those features generate the dominant contributions towards the higher-order moments.\\

\subsection{Electronic structure calculations}
 
The evaluation of vacuum polarization effects has been implemented within the relativistic electronic structure package \texttt{BERTHA}. It utilizes a finite basis of atom-centred Gaussian functions which satisfy the strict kinetic balance condition
\begin{equation}
	\mathcal{M}[S,\mu;\vect{r}] \propto (\vect{\sigma} \cdot \vect{p}) \mathcal{M}[L,\mu;\vect{r}].
\end{equation}
where $\mathcal{M}[T,\mu;\vect{r}]$ are two-component functions, $T=L$ denotes a large-component function, and $T=S$ a small-component function. The index $\mu$ serves as a label for the basis function, and is used as a shorthand notation for the relativistic quantum numbers $\kappa_\mu$ and $m_\mu$, a Gaussian exponent, $\lambda_{\mu}$, and the local origin of coordinates, $\vect{A}_{\mu}$. The kinetic balance condition enforces a one-to-one matching of large- and small-component basis functions.\\

The algorithms used to assemble the Fock matrix during this iterative self-consistent-field calculation~\cite{FlynnGrantQuiney2023} include one-centre contributions derived from Racah algebra, which offer particularly efficient integral algorithms. The electron-electron interaction includes both the Coulomb and the Breit interactions. This provides an effective potential approximation to the full one-photon exchange Feynman diagram between pairs of bound electrons in the Coulomb gauge in the low-frequency limit $\omega \to 0$.\\

\begin{figure}[!h]
	\begin{center}
		\includegraphics[width=8.5cm]{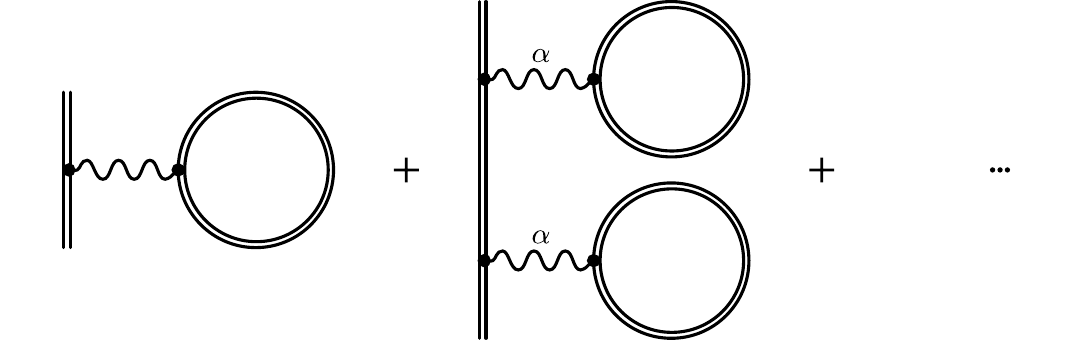}
		\caption{Including the order $\alpha$ potential within an electronic structure calculation sums an infinite series of independent vacuum polarization events.}
		\label{fig:vacuumpolarization_alpha_iterative}
	\end{center}
\end{figure}

The simultaneous inclusion of the relativistic four-component structure of the Dirac equation, detailed models of nuclear structure, the Coulomb and Breit interaction, and radiative vacuum polarization effects, define a new model, an effective QED hamiltonian. By including all these effects with the conventional Hartree-Fock mean-field model, we define a $\texttt{DHFQ}$ hamiltonian of the form
\begin{equation}
	\hat{\mathcal{H}}_{DCBQ} = \hat{\mathcal{H}}_{\texttt{DHFQ}} + \hat{\mathcal{H}}_{C}^{(1)} + \hat{\mathcal{H}}_{B}^{(1)}
\end{equation}
with
\begin{equation}
	\hat{\mathcal{H}}_{\texttt{DHFQ}} = \sum_{a=1}^{N_{e}} \Bigl( \hat{h}_{DQ}(\vect{r}_{a}) + u_{C}^{HF}(\vect{r}_{a}) + u_{B}^{HF}(\vect{r}_{a}) \Bigr).
\end{equation}
The first-order many-body correction energy due to the electron-electron interaction $\hat{u}_{ab} = \hat{g}_{ab} + \hat{b}_{ab}$,
\begin{equation}
	\mathcal{H}^{(1)} = \sum_{a=1}^{N_{e}} \sum_{b=a+1}^{N_{e}} \hat{u}_{ab} - \sum_{a=1}^{N_{e}} u^{HF} (\vect{r}_{a}),
\end{equation}
vanishes identically by Brillouin's theorem. In quantifying the nuclear, Uehling and Wichmann-Kroll potentials that arise for a heavy element in a Gaussian charge model, we construct the entire bound-state vacuum polarization diagram at order $\alpha$, to a good approximation. By incorporating that interaction within our mean-field procedure, we are able to account not only for the one-loop diagram, but also an infinite series of ladder diagrams at higher orders in $\alpha$, as indicated in Fig.~\ref{fig:vacuumpolarization_alpha_iterative}.

\subsection{ZCG fitting functions}
The vacuum polarization potentials induced by the presence of a finite nuclear charge density have been fitted to a set of auxiliary Gaussian functions which carry zero charge (ZCGs); we have described this procedure elsewhere~\cite{FlynnQuiney2024a}. These functions admit simple analytic matrix element evaluation for one-centre contributions by means of the integral
\begin{equation}
	\int_{0}^{\infty} r^{2 \ell+2} e^{-\lambda r^{2}} A_{n} e^{-\zeta_{n} r^{2}} \; \mathrm{d}r = \frac{A_{n}}{2} \frac{\Gamma ( \ell + 3/2 )}{(\lambda + \zeta_{n})^{\ell + 3/2}}.
\end{equation}
The appropriate fitting set for the Uehling interaction was reported previously~\cite{FlynnQuiney2024a}, and is used here. Wichmann-Kroll and K\"{a}ll\'{e}n-Sabry contributions from the electron-positron field as well as the leading-order contributions due to the charged pion and muon fields, have utilized ZCG fitting sets that may be found in the supplementary data.\\

For multi-centre matrix elements which involve the zero-charge Gaussian functions we utilize a finite expansion of the charge-current overlap densities in an auxiliary set of scalar Hermite functions of the form
\begin{eqnarray}
	\lefteqn{\mathcal{M}^{\dagger}[T,\mu;\vect{r}_{\mu}] \; \sigma_{q} \; \mathcal{M}[T',\nu;\vect{r}_{\nu}]}\\ \notag
	&=& \sum_{\alpha \beta \gamma} E_{q}^{TT'}[\mu,\nu;\alpha,\beta,\gamma] \mathcal{H}(p,\vect{r}-\vect{P};\alpha,\beta,\gamma),
\end{eqnarray}
where $p = \lambda_{\mu}+\lambda_{\nu}$ and the relative centre of the charge density is $\vect{P} = (\lambda_{\mu}\vect{C}_{\mu}+\lambda_{\nu}\vect{C}_{\nu})/p$. The Hermite basis is constructed from the product of one-dimensional Hermite-Gaussian functions,
\begin{equation}
	\mathcal{H}(p,\vect{r};\alpha,\beta,\gamma) = H_{\alpha}(p,x)H_{\beta}(p,y)H_{\gamma}(p,z),
\end{equation}
while the $E_{q}^{TT'}$-coefficients absorb normalization factors as well as the component-coupling details of a two-component basis function overlap.

\subsection{Atomic E119$^+$}

Zero-charge Gaussian density expansions facilitate the evaluation of multi-centre integrals in molecular systems. Taking a particular ZCG function to be defined with respect to a local origin of coordinates at $\vect{C}$, the vacuum polarization potential associated with that expansion generates a basis set matrix element 
\begin{eqnarray}\notag
	\widetilde{V}^{TT}_{\mu \nu}
	&=& \iiint \mathcal{M}^{\dagger}[T,\mu;\vect{r}_{\mu}] \; \sigma_{0} \widetilde{V}(\vect{r}-\vect{C}) \; \mathcal{M}[T,\nu;\vect{r}_{\nu}] \; \mathrm{d}^{3}\vect{r}\\ \notag
	&=& \sum_{\alpha \beta \gamma}^{\Lambda} E_{0}^{TT}[\mu,\nu;\alpha,\beta,\gamma] \iiint \mathcal{H}(p,\vect{r}-\vect{P};\alpha,\beta,\gamma) \\ \notag
	& & \hspace{0.5cm} \cdot \sum_{n=1}^{N} Z A_{n} \mathcal{H}(\zeta_{n},\vect{r}-\vect{C};0,0,0) \; \mathrm{d}^{3}\vect{r}\\
	&=& \sum_{\alpha \beta \gamma}^{\Lambda} E_{0}^{TT}[\mu,\nu;\alpha,\beta,\gamma] \\ \notag
	& & \hspace{0.5cm} \cdot Z \sum_{n=1}^{N} \bigl( \tfrac{\pi}{p+\zeta_{n}} \bigr)^{3/2} A_{n} \mathcal{H}\bigl(\alpha,\beta,\gamma;\tfrac{p \zeta_{n}}{p+\zeta_{n}},\vect{P}-\vect{C}\bigr).
\end{eqnarray}
This is equivalent in form to a three-centre density overlap matrix element, and requires very little computational effort to evaluate, particularly in comparison to a typical batch of Coulomb or Breit matrix elements. These matrix elements are easily evaluated for large- and small-component basis function products alike.

\section{Results and Discussion}\label{sec:results-discussion}
\subsection{Spatial variation in vacuum polarization potentials}
To gain some insight into the spatial variation of vacuum polarization potentials, we plot them in Fig.~\ref{fig:vp-potential-loglog} on a log-log scale. The nuclear potential due to a Gaussian charge distribution is included for comparison. The asymptotic behaviour of the nuclear Coulomb field is evidently of long-range. The vacuum polarization potentials that arise through the electron-positron field have a characteristic length $\lambdabar_e$, which is large compared to the nuclear radius, but small compared to the Bohr radius. In contrast, the potentials due to the muon and pion fields have a characteristic length of the order of one femtometre, and are strongly attentuated outside the nuclear volume.\\

For this atomic calculation of the ground-state of the E119$^{+}$ ion, we have selected a particularly large basis set involving a geometric series for all required angular types, using Gaussian parameter lists $75s 70p 65d 60f$. The parameters are generated by means of a geometric series $\lambda_{n} = \alpha \beta^{n-1}$ with $\alpha = 0.00845$ and $\beta = 1.55$. This large basis set serves as an appropriate benchmark to demonstrate the accuracy of basis set atomic calculations compared with finite difference representations. The basis functions with largest parameters $\lambda$ are sufficiently localized that they significantly penetrate the nucleus, providing the flexibility to represent the short-range character of vacuum polarization effects.\\

\begin{figure}[!t]
\begin{center}
\includegraphics[width=8.5cm]{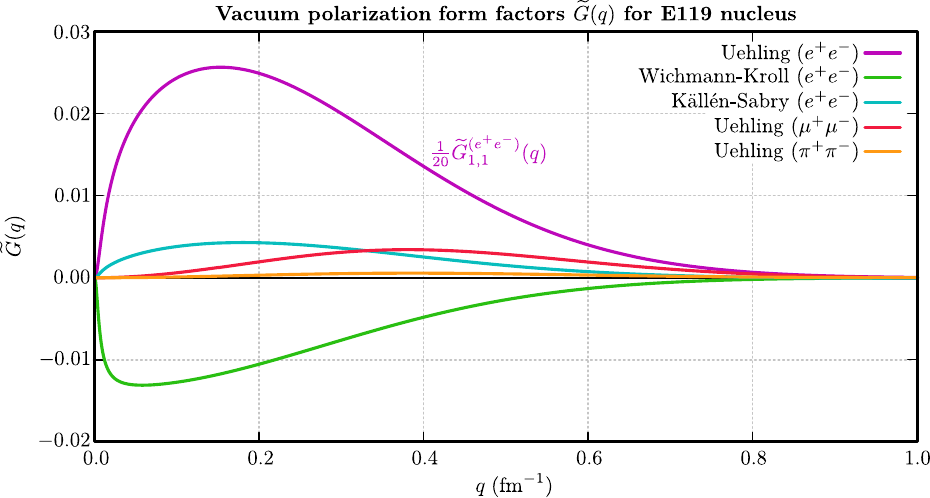}
\includegraphics[width=8.5cm]{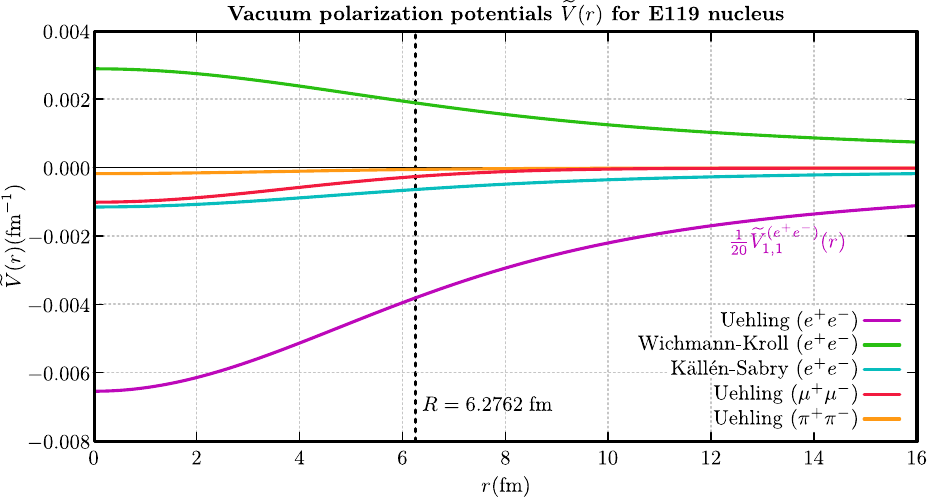}
\includegraphics[width=8.5cm]{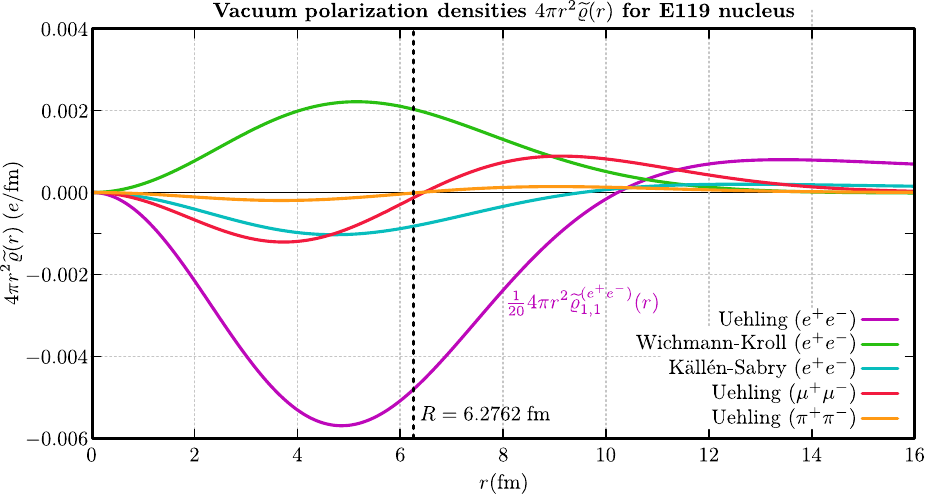}
\caption{Vacuum polarization interactions arising from higher-order contributions and additional virtual fields, for a E119 nucleus following a Gaussian charge model.}
\label{fig:vacpol_allfields}
\end{center}
\end{figure}

The vacuum polarization interactions which arise from an E119 nucleus, given a Gaussian nuclear charge model, are shown in Fig.~\ref{fig:vacpol_allfields}. This includes the form factors $\widetilde{G}(q)$ to represent the momentum space aspects of each interaction, the electrostatic potentials that they generate, and the radial charge distributions generated by each of those interactions.\\

\begin{table*}
    \caption{\label{tab:E119+_energies_methods}Energy contributions (a.u.) for the E119$^{+}$ ion, given a Gaussian nuclear charge model. The column marked \texttt{DHFR}$^{\ast}$ records vacuum polarization contributions evaluated using numerical quadrature; these quantities were evaluated using ZCG expansions in all other models. The \texttt{DHFR}$^{\ast}$ and \texttt{DHFR} models treat the Coulomb interaction self-consistently, and all other effects in first-order perturbation theory. The \texttt{DHFB} model treats the Coulomb and Breit interactions self-consistently, and all other effects in first-order perturbation theory. The \texttt{DHFQ} model includes all effects self-consistently.}
    \begin{ruledtabular}
    \begin{tabular}{ l | r | r r r }
                                        & \texttt{DHFR}$^{\ast}$ & \texttt{DHFR} & \texttt{DHFB} &    \texttt{DHFQ} \\ \hline
        Dirac                           & --74814.85434779 & --74814.85434779 & --74785.30449729 & --74788.52257843 \\
        Coulomb                         &   18669.55418002 &   18669.55418002 &   18640.22401196 &   18643.48811978 \\
        Breit                           &     102.63126974 &     102.63126974 &     102.19144558 &     102.43298257 \\
        Uehling ($e^{+}e^{-}$)          &    --50.20593583 &    --50.20593567 &    --49.95173100 &    --50.54743968 \\ 
        Wichmann-Kroll ($e^{+}e^{-}$)   &       3.17357393 &       3.17367986 &       3.15825525 &       3.18577222 \\ 
        K\'{a}llen-Sabry ($e^{+}e^{-}$) &     --0.37422074 &     --0.37421989 &     --0.37233466 &     --0.37674178 \\ 
        Uehling ($\mu^{+}\mu^{-}$)      &     --0.01569987 &     --0.01570055 &     --0.01561843 &     --0.01599158 \\ 
        Uehling ($\pi^{+}\pi^{-}$)      &     --0.00243145 &     --0.00243055 &     --0.00241784 &     --0.00247572 \\ \hline
        Total                           & --56090.09360849 & --56090.09350483 & --56090.07288643 & --56090.35835262 \\  
    \end{tabular}
    \end{ruledtabular}
\end{table*}

Table~\ref{tab:E119+_energies_methods} summarizes calculations made on the E119$^+$ ion using several mean-field hamiltonians and computational methods within \texttt{BERTHA}. In the column marked $\texttt{DHFR}^{\ast}$, electronic spinor amplitudes have been sampled on a radial grid which consists of 20,000 equally-spaced points on the interval $[0,10 \text{ fm}]$ and a further 180,000 exponentially-spaced points on $[10 \text{ fm},35 \text{ a.u.}]$. These amplitudes are obtained from self-consistent solutions of the Dirac-Hartree-Fock-Roothaan equation within the finite basis set approach. They are combined with the appropriate vacuum polarization potentials, which have been calculated using our convolution-based approach in the vicinity of the nuclear distribution, and the Taylor series evaluation in Eq.~(\ref{eq:vp-fr-potential}) elsewhere. The radial integrations are carried out with an 11-point Newton-Cotes quadrature scheme. These results are consistent with those of our ZCG expansions -- almost all expectation values are consistent with their counterparts evaluated on a radial grid to within the $\mu$ha level. The only exception to this result is for the Wichmann-Kroll calculation, which involves a factor of $Z^{3}$. For E119, this is $Z^{2} = 14,161$ times greater in amplitude than the first-order diagrams. This behaviour cannot be subsumed by our ZCG expansions, and so we must content ourselves with an expansion that is accurate only to the 0.1~mha level for this superheavy species.

\begin{widetext}
	\begin{center}
		\begin{figure}[!b]
			\includegraphics[width=15.0cm]{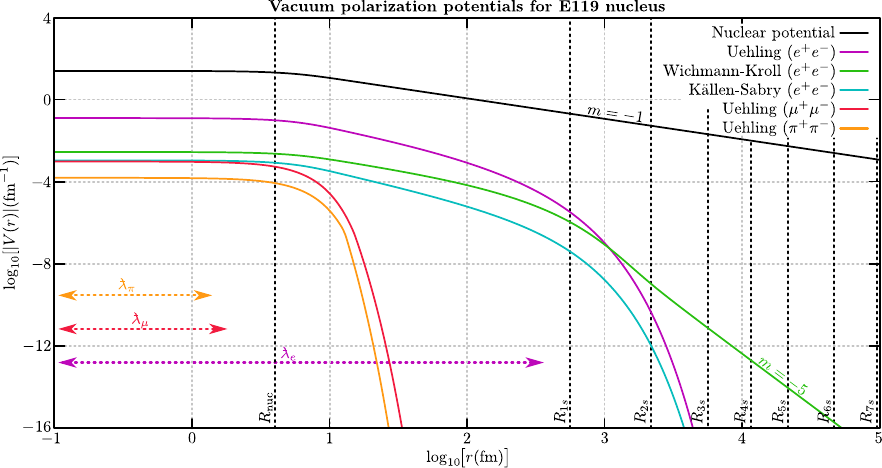}
			\caption{Vacuum polarization potentials generated by a Gaussian nuclear distribution with $Z=119$ and $R = 6.2762$ fm.\\ The radial influence of these interactions on atomic electrons is indicated by marking the spatial extent of their charge densities.}
			\label{fig:vp-potential-loglog}
		\end{figure}
	\end{center}
\end{widetext}

\subsection{Molecular E119F}\label{sec:E119F_molecule}

Using the molecular module of the \texttt{BERTHA} package, we have performed a calculation for the ground state of the E119F molecule using an equilibrium bond length of $d = 2.448 \angstrom$, for which results are listed in Table~\ref{tab:E119F_breakdown}. In this approach, we have used Racah-based algorithms to handle the one-centre Coulomb and Breit integrals, which circumvents a severe computational bottleneck where the high-$\ell$ basis functions on the E119 centre are concerned.\\

The ZCG expansions of the effective vacuum polarization interactions for the E119$^{+}$ and F$^-$ ions may now be applied to the \texttt{DHFQ} mean-field hamiltonian for the closed-shell diatomic molecule. Having previously discussed the challenges of managing the relativistic Coulomb and Breit interactions in a molecular context, it suffices here simply to state that both can be handled within a reasonable computational budget, provided that a somewhat more modest basis set is adopted for the molecular problem. For the E119 atomic centre we have selected the $34s31p23d18f$ basis set of Miranda \text{et al.}~\cite{miranda2012} which possesses Fock matrix dimension $N_B=1472$, while the atomic fluorine centre for this molecule involves Dunning's optimized aug-cc-pVTZ basis set parameters~\cite{dunning1989}, with a resulting $18s 10p 3d$ list and a Fock matrix extension of $N_B=244$. The charge characteristics of the E119 nucleus are identical to our previous calculation, while for the fluorine centre we have used a Gaussian charge model with $A=18.99$ and a corresponding charge radius $R = 2.8004$ fm.

\begin{table*}
    \caption{\label{tab:E119F_breakdown} Energy expectation values (a.u.) involving the \texttt{DHFQ} hamiltonian for the E119F molecule, using basis sets from Miranda \text{et al.}~\cite{miranda2012} and Dunning~\cite{dunning1989} for E119$^{+}$ and F$^{-}$ respectively. \textsc{Ionic fragments}: The ionic constituents of the E119F molecule, generated from atomic basis set calculations. \textsc{Molecular contributions}: Decomposition into one- and multi-centre contributions from a molecular calculation, using a bond length $d = 2.448 \angstrom$. The Breit and vacuum polarization interactions lead to multi-centre contributions that are small compared to one-centre terms. ``WK'' refers to the Wichmann-Kroll interaction and ``KS'' refers to the K\"all\'en-Sabry interaction.}
    \begin{ruledtabular}
    \begin{tabular}{ l | r r | r || r r | r}
        & \multicolumn{3}{c||}{\textsc{Ionic fragments}} & \multicolumn{3}{c}{\textsc{Molecular contributions}}\\ \hline
        &       E119$^{+}$ &        F$^{-}$ &     Fragment sum & One-centre       & Multi-centre   & Molecule         \\ \hline
        Nuclei                     &               -- &             -- &               -- &               -- &   231.51502982 &     231.51502982 \\
        Dirac                      & --74788.51401150 & --144.11974286 & --74932.63375436 & --74953.93753780 & --464.33876344 & --75418.27630124 \\
        Coulomb                    &   18643.48032218 &    44.57475210 &   18688.05507428 &   18708.86569473 &   233.10509011 &   18941.97078484 \\
        Breit                      &     102.43298397 &     0.01140572 &     102.44438969 &     102.44438603 &     0.00010738 &     102.44449340 \\
        Uehling ($e^{+}e^{-}$)     &    --50.54748132 &   --0.00038832 &    --50.54786964 &    --50.54761497 &   --0.00009104 &    --50.54770602 \\
        WK ($e^{+}e^{-}$)          &       3.18577269 &     0.00000033 &       3.18577302 &       3.18576641 &     0.00000516 &       3.18577158 \\
        KS ($e^{+}e^{-}$)          &     --0.37674221 &   --0.00000338 &     --0.37674559 &     --0.37674361 &   --0.00000067 &     --0.37674428 \\
        Uehling ($\mu^{+}\mu^{-}$) &     --0.01599154 &   --0.00000001 &     --0.01599155 &     --0.01599106 &   --0.00000003 &     --0.01599109 \\
        Uehling ($\pi^{+}\pi^{-}$) &     --0.00247570 &   --0.00000000 &     --0.00247570 &     --0.00247563 &   --0.00000001 &     --0.00247563 \\ \hline
        Total                      & --56090.35762343 &  --99.53397642 & --56189.89159985  & --56190.38451590 &     0.28137728 & --56190.10313862 \\
        \end{tabular}
    \end{ruledtabular}
\end{table*}

\section{Conclusion}\label{sec:conclusion}

A formalism has been outlined which demonstrates common features of vacuum polarization interactions in effective QED. The redistribution of virtual pairs due to the presence of a source charge is investigated from the perspectives of form factors in momentum space, as well as electrostatic potentials and charge densities in coordinate space. A general strategy for generating vacuum polarization potentials which arise from detailed nuclear charges has been devised, and implemented for a number of such interactions which arise from an E119 nucleus. A fitting scheme for representing those potentials has been employed, and applied towards the E119F molecule in a relativistic mean-field calculation.\\

The self-consistent treatment of the Breit and effective QED interactions greatly facilitates the subsequent incorporation of these effects in many-body electronic structure calculations. Taken at face value, such a strategy carries a significant computational cost. As we have demonstrated in this and elsewhere~\cite{FlynnQuiney2024a,FlynnGrantQuiney2023}, however, these effects are so strongly localized to the spatial regions around the nuclear charge distributions that, for all practical purposes, they may be treated accurately using the one-centre methods of atomic physics, with a scaling that is linearly proportional to the number of nuclear centres. The error made in neglecting multi-centre effects is less than one part in a million, in energy contributions that are themselves small corrections to the dominant Dirac-Coulomb hamiltonian. Moreover, the self-consistent inclusion of these effects modifies single-electron Dirac four-spinors in the neighbourhood of the nuclei. They provide, therefore, a more complete representation of the physics in the evaluation of properties that are sensitive to the details of the electronic amplitudes in this region. These effects include hyperfine interactions, Standard Model interactions that are odd with respect to parity and time reversal symmetries, as well as interactions that arise in speculative theories of particle physics beyond the Standard Model. The use of the theoretical and mathematical methods of atomic and molecular physics to test the Standard Model, and to probe theories designed to overcome its limitations, remains a strong motivation for the continued development of this work.

\bibliographystyle{apsrev4-1}
\bibliography{VP_II}

\end{document}